\PassOptionsToPackage{unicode}{hyperref}
\PassOptionsToPackage{hyphens}{url}
\documentclass[12pt]{article}
\usepackage{amsmath}
\usepackage{graphicx,psfrag,epsf}
\usepackage{enumerate}
\usepackage[]{natbib}
\usepackage{textcomp}

\newcommand{\blind}{0}

\usepackage{amsmath,amssymb,mathtools,bbm}
\usepackage{tikz}
\usetikzlibrary{positioning, shapes, intersections, through, backgrounds, fit, decorations.pathmorphing}

\usepackage{algorithm}
\usepackage{algpseudocode}

\usepackage{setspace}
\usepackage{pdflscape}
\usepackage{placeins}

\usepackage{booktabs}
\usepackage{colortbl}

\usepackage{color}
\definecolor{myredhighlight}{RGB}{180, 15, 32}
\definecolor{mydarkblue}{RGB}{0, 33, 79}
\definecolor{mymidblue}{RGB}{44, 127, 184}
\definecolor{mylightblue}{RGB}{166, 233, 255}
\definecolor{mywhwlow}{RGB}{234, 164, 99}

\usepackage{accents}
\newlength{\dhatheight}

\newcommand{\pd}{\text{p}}
\newcommand{\Pd}{\text{P}}
\newcommand{\q}{\text{q}}
\newcommand{\Q}{\text{Q}}

\newcommand{\tc}{\text{T}}
\newcommand{\tp}{\text{t}}

\DeclarePairedDelimiterX{\infdivx}[2]{(}{)}{%
  #1\;\delimsize\|\;#2%
}

\newcommand{\kldiv}{\text{KL}\infdivx}

\algrenewcommand\alglinenumber[1]{
    {\sf\footnotesize\color{lightgray}#1}}
\algrenewcommand\algorithmicrequire{\textbf{Inputs:}}
\algrenewcommand\algorithmicensure{\textbf{Postcondition:}}

\def\dodoi#1{doi: \href{https://doi.org/#1}{\nolinkurl{#1}}}
\def\dourl#1{\href{http://#1}{\nolinkurl{#1}}}

\IfFileExists{bookmark.sty}{\usepackage{bookmark}}{\usepackage{hyperref}}
\IfFileExists{xurl.sty}{\usepackage{xurl}}{} 
\hypersetup{
  pdftitle={Translating predictive distributions into informative priors},
  pdfkeywords={prior specification, multi-objective optimisation, prior
predictive checks},
  hidelinks,
  pdfcreator={LaTeX via pandoc}}

\begin{document}

\def\spacingset#1{\renewcommand{\baselinestretch}%
{#1}\small\normalsize} \spacingset{1}


\if0\blind
{
  \title{\bf Translating predictive distributions into informative
priors}

  \author{
        Andrew A. Manderson \thanks{a.manderson@live.co.uk} \\
    MRC Biostatistics Unit, University of Cambridge\\
     and \\     Robert J. B. Goudie \\
    MRC Biostatistics Unit, University of Cambridge\\
      }
  \maketitle
} \fi

\if1\blind
{
  \bigskip
  \bigskip
  \bigskip
  \begin{center}
    {\LARGE\bf Translating predictive distributions into informative
priors}
  \end{center}
  \medskip
} \fi

\bigskip
\begin{abstract}
``When complex Bayesian models exhibit implausible behaviour, one
solution is to assemble available information into an informative prior.
Challenges arise as prior information is often only available for the
observable quantity, or some model-derived marginal quantity, rather
than directly pertaining to the (usually latent) parameters in our
model. We propose a method for translating available prior information,
in the form of an elicited distribution for the observable or
model-derived marginal quantity, into an informative joint prior. Our
approach proceeds given a parametric class of prior distributions with
as yet undetermined hyperparameters, and minimises the difference
between the supplied elicited distribution and corresponding prior
predictive distribution. We employ a global, multi-stage Bayesian
optimisation procedure to locate optimal values for the hyperparameters.
Three examples illustrate our approach: a cure-fraction survival model,
where censoring implies that the observable quantity is \emph{a priori}
a mixed discrete/continuous quantity; a setting in which prior
information pertains to \(R^{2}\) -- a model-derived quantity; and a
nonlinear regression model.''
\end{abstract}

\noindent%

\vfill

\newpage
\spacingset{1.9} 

\hypertarget{introduction}{%
\section{Introduction}\label{introduction}}

Incorporating prior information in Bayesian models is conceptually easy,
but in practice constructing an informative prior is not easy.
Formulating priors in accordance with predictive information obtained
via predictive elicitation \citep{ohagan_uncertain_2006} is attractive
due to the widespread availability, reliability
\citep{kadane_experiences_1998} and model-agnostic nature of such
information. However it is often unclear how to implement this approach,
particularly for complex, nonlinear, or overparameterised models, for
which informative priors can be essential to exclude model behaviours
that conflict with known properties of the world. In this paper we
suppose predictive information is available in the form of a
\emph{target} prior predictive distribution, and consider how to
\emph{translate} this into a prior distribution for model parameters, a
step that has heretofore received relatively little attention.

One simple approach to this task is to directly model the elicited
quantity. This requires no translation step. For example,
\citet{perepolkin_hybrid_2021} directly updated elicited information in
light of observations using a Bayesian quantile-parameterised
likelihood. Such direct approaches are currently only feasible for
simple models with no latent structure. For models with simple latent
structure, eliciting information about an invertible function of the
parameters may be possible \citep[e.g.][]{chaloner_graphical_1993},
enabling analytic translation into a prior for the parameters.
Translation is also clear for conjugate distributions, since the prior
predictive distribution determines the prior hyperparameter values
\citep{percy_bayesian_2002}. Translation, however, is unclear in general
for nonconjugate models \citep{gribok_backward_2004}. Techniques exist
for specific models with specific latent structures, including for
logistic regression \citep{chen_prior_1999}, contingency table analyses
\citep{good_bayesian_1967} and hierarchical models
\citep{hem_robustifying_2021}, but a model-agnostic approach is needed
for models outwith these classes, as noted by
\citet{mikkola_Prior_2023}.

Our approach to translation builds on the idea of predictive checks
\citetext{\citealp{gabry_visualization_2019}; \citealp[the
``hypothetical future samples'' of][]{winkler_assessment_1967}} and the
Bayesian workflow \citep{gelman_bayesian_2020}, in which the prior is
repeatedly adjusted until there is concordance between the prior
predictive distribution and the elicited predictive information.
However, this manual approach is impractical whenever the relationship
between the prior and the distribution of the observables is muddied by
the complexity of the intervening model. A more automated method is
required. \citet{wang_using_2018} and \citet{thomas_probabilistic_2020}
have proposed approaches in which either regions of observable space or
specific realisations are labelled as plausible or implausible by
experts, and then a prior accounting for this information is formed via
either history matching or a ``human in the loop'' process driven by a
Gaussian process model. \citet{albert_combining_2012} propose a
supra-Bayesian approach intended for multiple experts, in which a
Bayesian model is formed for quantiles or probabilities elicited from
the experts. Another approach, and the closest in motivation and
methodology to ours, is \citeauthor{hartmann_flexible_2020}
\citetext{\citeyear{hartmann_flexible_2020}; \citealp[which is partly
inspired by][]{da_silva_prior_2019}}, which employs a Dirichlet
likelihood for elicited predictive quantiles to handle both elicitation
and translation. Our approach is model-agnostic and is fully based
around distributions, meaning uncertainty is directly and intuitively
represented. We specify a suitable, generic loss function between the
prior predictive distribution and this target distribution, and minimise
this loss function via a generic, multi-objective global optimisation
process. We implement our methodology in an \texttt{R} package
\texttt{pbbo} (\url{https://github.com/hhau/pbbo}; Supplement
\ref{r-package}).

\hypertarget{methodology}{%
\section{Methodology}\label{methodology}}

We postulate three properties that we would like our method to satisfy:

\emph{Faithfulness}~~ A prior is faithful if it accurately encodes the
target data distribution provided by the elicitation subject.
Faithfulness is a property of both the procedure employed to obtain the
prior and the model itself, since not all target prior predictive
distributions can be encoded by simple models and prior structures.
Faithfulness is related to the definition of \emph{validity} in
\citet{johnson_valid_2010} and \citet{ohagan_uncertain_2006}'s use of
\emph{faithful}, but their concerns are specific to the elicitation
process and not to the translational step.

\emph{Uniqueness}~~ Multiple equally faithful prior distributions may
exist in complex models, meaning we must distinguish between such priors
based on properties other than just faithfulness if a unique prior is
desired. For example, if maximising prior uncertainty whilst retaining
faithfulness is desired, then we could choose the prior with the largest
marginal standard deviations (see Section
\ref{regularising-estimates-of-lambda-secondary-objective}). Other
properties could be used similarly. The challenge of uniqueness has been
noted by \citet{stefan_practical_2022}.

\emph{Replicability}~~ A procedure is replicable if, given the same
target, it constructs identical priors across independent replications.
This is unlikely to hold exactly with stochastic algorithms, meaning it
is important to assess.

\hypertarget{setup}{%
\subsection{Setup}\label{setup}}

Consider a joint probability distribution for an observable
\(Y \in \mathcal{Y} \subseteq{} \mathbb{R}\) and parameters
\(\theta \in \Theta \subseteq \mathbb{R}^{Q}\), given hyperparameters
\(\lambda \in \Lambda \subset \mathbb{R}^{L}\). This distribution has
cumulative distribution function (CDF) \(\Pd(Y, \theta \mid \lambda)\)
and prior predictive CDF \(\Pd(Y \mid \lambda)\) for \(Y\). We suppose a
target predictive distribution, with CDF \(\tc(Y)\), for the observable
quantity \(Y\) has been chosen, and that this encapsulates our prior
knowledge about \(Y\). Our primary aim is to choose \(\lambda\) so that
the prior predictive \(\Pd(Y \mid \lambda)\) is faithful to the target
\(\tc(Y)\).

We assume that the target \(\tc(Y)\) can be described by a (mixture of)
standard distributions and that samples can be drawn from it; but we do
not require \(\tc(Y)\) to be in the same parametric family as the prior
predictive \(\Pd(Y \mid \lambda)\), since this is often unavailable in
closed-form. We recommend choosing \(\tc(Y)\) using predictive
elicitation \citep{kadane_experiences_1998}, in which an appropriate
parametric distribution is fitted to a small number of quantiles (of the
observable quantity) elicited from experts
\citep[chap.~6]{ohagan_uncertain_2006}.

We describe our methodology in a slightly more general setting in which
the observable quantity \(Y\) is conditional on a covariate
\(X \in \mathcal{X} \subseteq \mathbb{R}^{C}\), with joint probability
distribution CDF \(\Pd(Y, \theta \mid \lambda, X)\) and prior predictive
CDF \(\Pd(Y \mid \lambda, X)\). We assume information has been elicited
about \(Y\) at a fixed set of values for \(X\). Specifically we suppose
the target CDF \(\tc(Y \mid X_{r})\) has been elicited at \(R\) values
of the covariate vector denoted \(\{X_{r}\}_{r = 1}^{R}\), which we
stack in the matrix
\(\boldsymbol{X} = \left[X_{1}^{\top} \cdots X_{R}^{\top}\right] \in \boldsymbol{\mathcal{X}} \subseteq \mathbb{R}^{R \times C}\).
We assume that each target \(\tc(Y \mid X_{r})\) has identical support
to \(\Pd(Y \mid \lambda, X_{r})\). We denote
\(\tc(Y \mid \boldsymbol{X}) = \prod_{r = 1}^{R} \tc(Y \mid X_{r})\),
with \(\Pd(Y \mid \lambda, \boldsymbol{X})\) and
\(\Pd(\theta \mid \lambda, \boldsymbol{X})\) defined analogously.

\hypertarget{predictive-discrepancy-primary-objective}{%
\subsection{Predictive discrepancy (primary
objective)}\label{predictive-discrepancy-primary-objective}}

We quantify the difference between the prior predictive and target by
the \emph{covariate-specific predictive discrepancy}, which we define to
be
\begin{equation}
  \tilde{D}(\lambda \mid \boldsymbol{X}) =
    \frac{1}{R}
    \sum_{r = 1}^{R}
    \int d(
      \Pd(Y \mid \lambda, X_{r}),
      \tc(Y \mid X_{r})
    ) \text{d}\tc(Y \mid X_{r}),
  \label{eqn:theoretical-discrep-definition-covariate}
\end{equation}\noindent
for some discrepancy function \(d(\cdot, \cdot)\). Minimising
\eqref{eqn:theoretical-discrep-definition-covariate} admits the optimal
hyperparameter
\(\lambda^{*} = \min_{\lambda \in \Lambda} \tilde{D}(\lambda \mid \boldsymbol{X})\).
The covariate-independent equivalent \(\tilde{D}(\lambda)\) is obtained
by setting \(R = 1\) and ignoring conditioning on \(X_{r}\).

Many forms of discrepancy function \(d(\cdot, \cdot)\) could be adopted,
but restricting to proper scoring rules \citep{gneiting_strictly_2007},
which are minimised iff
\(\Pd(Y \mid \lambda, X_{r}) = \tc(Y \mid \lambda)\) for all
\(Y \in \mathcal{Y}\), is intuitive. In this case, if
\(\Pd(Y \mid \lambda, X_{r})\) is flexible enough to exactly match
\(\tc(Y \mid X_{r})\) for some \(\lambda^{*}\), then any such
discrepancy will yield the same \(\lambda^{*}\). Differences arise when
\(\Pd(Y \mid \lambda, X_{r})\) is insufficiently flexible, with
discrepancy functions differing in placement of emphasis.

CDF-based discrepancies are appealing because they are
widely-applicable, so inspired by the Cramér-von Mises
\citep{von_mises_asymptotic_1947} and Anderson-Darling
\citep{anderson_asymptotic_1952} distributional tests we define, for
CDFs \(\text{M}(Y)\) and \(\Pd(Y)\):
\begin{equation}
  d^{\text{CvM}}(\text{M}(Y), \Pd(Y)) = \left(\text{M}(Y) - \Pd(Y\right))^{2}, \quad
  d^{\text{AD}}(\text{M}(Y), \Pd(Y)) = \frac{
    \left(\text{M}(Y) - \Pd(Y\right))^{2}
  } {
    \Pd(Y) (1 - \Pd(Y))
  }.
  \label{eqn:discrepancies-definitions}
\end{equation}\noindent
The Anderson-Darling (AD) discrepancy \(d^{\text{AD}}\) places more
emphasis than Cramér-von Mises (CvM) on matching the tails of two CDFs.

Another option is either direction of Kullback-Leibler (KL) divergence,
\begin{equation}
  d^{\text{KL-fwd}} = \kldiv{\text{M}(Y)}{\Pd(Y)}, \quad
  d^{\text{KL-rev}} = \kldiv{\Pd(Y)}{\text{M}(Y)}.
  \label{eqn:forward-reverse-kl-defs}
\end{equation}
\noindent
We consider the form of KL divergences detailed in Supplement
\ref{using-the-kullbackleibler-divergence-as-a-discrepancy}.

\hypertarget{regularising-estimates-of-lambda-secondary-objective}{%
\subsection{\texorpdfstring{Regularising estimates of \(\lambda^{*}\)
(secondary
objective)}{Regularising estimates of \textbackslash lambda\^{}\{*\} (secondary objective)}}\label{regularising-estimates-of-lambda-secondary-objective}}

There often are many optimal values \(\lambda^{*}\) that yield values of
\(\tilde{D}(\lambda^{*} \mid \boldsymbol{X})\) that are practically
indistinguishable \citep[noted by][]{da_silva_prior_2019} but with
immensly differing prior distributions
\(\Pd(\theta \mid \lambda^{*}, \boldsymbol{X})\). That is, there are
many equally faithful priors. This is not surprising because we are
providing information only on \(Y\), which is typically of lower
dimension than \(\theta\). A particularly challenging case for
uniqueness is in models with additive noise forms, such as
\eqref{eqn:preece-baines-model-definition-one}; in this case it will
generally be necessary to fix a prior for the noise using knowledge of
the measurement process.

To handle more general cases of lack of uniqueness, we define a
secondary objective \(\tilde{N}(\lambda \mid \boldsymbol{X})\),
typically via a function \(n(\theta)\) with
\begin{equation}
  \tilde{N}(\lambda \mid \boldsymbol{X}) = \int n(\theta) \, \text{d} \Pd(\theta \mid \lambda, \boldsymbol{X}).
  \label{eqn:generic-secondary-objective}
\end{equation}\noindent
This objective can be chosen by practitioners to promote or inhibit
properties of the prior predictive \(\Pd(Y \mid \lambda, X_{r})\) as
desired.

As an example, we demonstrate how to encode a preference for maximising
prior uncertainty, as is commonly desired in the absence of contrary
prior knowledge. Specifically, given two estimates of \(\lambda^{*}\)
which have equivalent values of
\(\tilde{D}(\lambda^{*} \mid \boldsymbol{X})\), we prefer the one with
the larger variance for
\(\Pd(\theta \mid \lambda^{*}, \boldsymbol{X})\). This preference could
be encoded in several ways: a simple option is the (negative) mean of
the marginal log standard deviations across the \(Q\) components of
\(\theta \in \Theta \subseteq \mathbb{R}^{Q}\).
\begin{equation}
  \tilde{N}(\lambda \mid \boldsymbol{X}) = -\frac{1}{Q}\sum_{q = 1}^{Q}
    \log\left(
      \text{SD}_{\,\Pd(\theta_{q} \mid \lambda, \boldsymbol{X})}\left[\theta_{q}\right]
    \right),
  \label{eqn:second-objective-def}
\end{equation}\noindent
where \(\text{SD}_{\Pd(Z)}[Z]\) is the standard deviation of \(Z\) under
distribution \(\Pd(Z)\). Analytic expressions for
\(\text{SD}_{\Pd(\theta \mid \lambda, \boldsymbol{X})}[\theta_{q}]\) can
be used if available; or Monte Carlo estimates otherwise.

\hypertarget{algorithm-and-optimisation}{%
\subsection{Algorithm and
optimisation}\label{algorithm-and-optimisation}}

We jointly minimise \eqref{eqn:theoretical-discrep-definition-covariate}
and \eqref{eqn:second-objective-def} using a multi-objective
optimisation algorithm, and obtain a set of possible \(\lambda\) values
which comprise the Pareto frontier
\(\mathcal{P} = \{\lambda_{l}\}_{l = 1}^{\lvert \mathcal{P} \rvert}\).
This is the set of all ``non-dominated'' choices for \(\lambda\),
meaning that no point in \(\mathcal{P}\) is preferable in \emph{both}
objectives to any of the remaining points in \(\mathcal{P}\)
\citep[chap.~2]{deb_multi-objective_2001}. For each \(\lambda\) in
\(\mathcal{P}\) we compute the loss
\begin{equation}
  \tilde{L}(\lambda) = \log(\tilde{D}(\lambda \mid \boldsymbol{X})) + \kappa \, \tilde{N}(\lambda \mid \boldsymbol{X}),
  \label{eqn:loss-definition}
\end{equation}\noindent
where the value of \(\kappa > 0\) expresses our relative belief in the
importance of the secondary objective. The optimal value is then
\(\lambda^{*} := \min\limits_{\lambda \in \mathcal{P}} \tilde{L}(\lambda)\).

This optimum depends on \(\kappa\), which will usually be difficult to
assess. However, using multi-objective optimisation we can evaluate
\eqref{eqn:loss-definition} for any \(\kappa\) without needing to redo
the optimisation step, and thus plot Pareto frontiers for a wide range
of values \(\kappa \in \mathcal{K}\) coloured by loss, with the minimum
loss point indicated. These can guide our choice of \(\kappa\): we can
seek a value of \(\kappa\) with the minimum loss point not on the
extreme of the Pareto frontier, since we would like to balance the two
objectives. This approach is particularly useful in settings where the
scales of the two optima differ markedly, which we further discuss in
Supplement \ref{further-notes-on-choosing-kappa}. Where it is feasible
to replicate the optimisation procedure, we can additionally seek a
choice of \(\kappa\) that leads to Pareto frontiers with minimal
variability across replicates, since this suggests the optimal solution
can be estimated reliably.

We use a two-stage global optimisation process. Our algorithm requires:
a method for sampling \(\Pd(Y \mid \lambda, \boldsymbol{X})\); upper and
lower limits that render \(\Lambda\) a compact subset of
\(\mathbb{R}^{L}\), due to our use of global optimisation; and methods
to evaluate the log-target CDF \(\log(\tc(Y \mid \boldsymbol{X}))\) and
for drawing samples according to \(\tc(Y \mid \boldsymbol{X})\). The
first optimisation stage in our algorithm considers only
\(\tilde{D}(\lambda \mid \boldsymbol{X})\) to focus on faithfulness,
whereas the second stage also considers
\(\tilde{N}(\lambda \mid \boldsymbol{X})\) to account for uniqueness and
replicability. We adopt this approach because minimising
\(\tilde{D}(\lambda \mid \boldsymbol{X})\) is considerably more
challenging than minimising \(\tilde{N}(\lambda \mid \boldsymbol{X})\).
An idealised form of this process is illustrated in Figure
\ref{fig:idealised_process}. We briefly describe the algorithm below;
full details are in Supplement \ref{algorithm-and-optimisation-details}.

\begin{figure}

{\centering \includegraphics{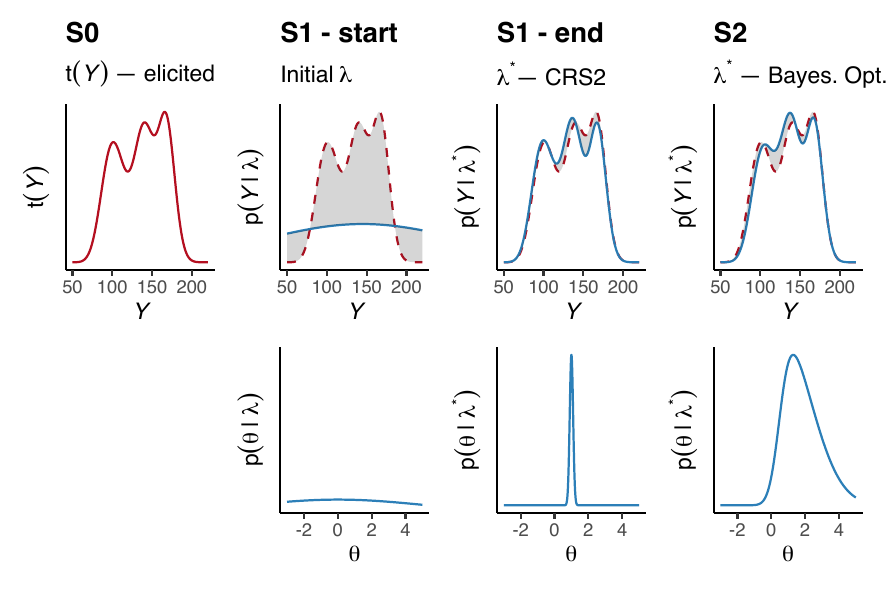} 

}

\caption{Illustration of the algorithm, which seeks to match the prior (\textcolor{mymidblue}{blue}) and target distribution (\textcolor{myredhighlight}{red}) by optimising $\lambda$. The initial value $\lambda$ produces a poor match. Stage 1 minimises \eqref{eqn:theoretical-discrep-definition-covariate}; stage 2 then minimises \eqref{eqn:loss-definition}, which increases the variance of $p(\theta \mid \lambda^{*})$.}\label{fig:idealised_process}
\end{figure}

In stage one we minimise \(\tilde{D}(\lambda \mid \boldsymbol{X})\)
using controlled random search 2 (CRS2) with local mutation
\citep{kaelo_variants_2006}, which we run for \(N_{\text{CRS2}}\)
iterations. We use the final optimum value \(\lambda^{*}\), as well as
the \(N_{\text{CRS2}}\) trial points, to obtain a design \(\mathcal{D}\)
for the next stage. The design comprises values of \(\lambda\), and
their corresponding values of
\(\log(\tilde{D}(\lambda \mid \boldsymbol{X}))\). A (small) number of
padding points \(N_{\text{pad}}\) are added to \(\mathcal{D}\) for
numerical robustness in stage 2. The result is the design
\(\mathcal{D} = \left\{\lambda_{i}, \log(\tilde{D}(\lambda_{i} \mid \boldsymbol{X}))\right\}_{i = 1}^{N_{\text{design}} + N_{\text{pad}}}\).
Whilst CRS2 was not designed to minimise noisy functions, empirically it
appears robust to small quantities of noise.

Stage one output is then used to initialise stage two, which
additionally focuses on uniqueness and replicability by employing
multi-objective Bayesian optimisation \citep{frazier_tutorial_2018} via
MSPOT \citep{zaefferer_mspot_2012} to jointly minimise
\(\tilde{D}(\lambda \mid \boldsymbol{X})\) and
\(\tilde{N}(\lambda \mid \boldsymbol{X})\). MSPOT uses a separate
Gaussian process (GP) approximation to each of the objectives, and
evaluates these approximations at many points from a Latin hypercube
design. At each iteration the best points under the current GP
approximations are evaluated using the actual objectives and used to
iteratively improve the approximations. The noisy (in practice) and
computationally expensive nature of our objectives, particularly
\(\tilde{D}(\lambda \mid \boldsymbol{X})\), necessitates an approach
such as MSPOT. Employing GP models for the objectives enables
inexpensive screening of values of \(\lambda \in \Lambda\) that are far
from optimal. Moreover, the GP is a flexible yet data efficient model to
use as an approximation and can, through appropriate choice of kernel,
capture correlation or other complex relationships between components of
\(\lambda\) and the objective. We use an optional batching technique in
stage two because the computational cost of evaluating the GP grows
cubically in the number of points \(N_{\text{BO}}\) used in its
construction. After \(N_{\text{BO}}\) iterations, the evaluated points
are reduced to their Pareto frontier \citep{kung_finding_1975}. Note
finding the global optimium is not guaranteed by our optimisation
strategy \citep{mullen_continuous_2014}.

To approximate \(\tilde{D}(\lambda \mid \boldsymbol{X})\) we first
approximate the prior predictive CDF
\(\Pd(Y \mid \lambda, \boldsymbol{X})\) by drawing \(S_{r}\) samples
\(\boldsymbol{y}_{r}^{(\Pd)} = (y_{s, r})_{s = 1}^{S_{r}}\) with
\(\boldsymbol{y}_{r}^{(\Pd)} \sim \Pd(Y \mid \lambda, X_{r})\) to form
the ECDF
\(\hat{\Pd}(Y \mid \lambda, X_{r}, \boldsymbol{y}_{r}^{(\Pd)})\), given
values of \(\lambda\) and \(X_{r}\). We then approximate
\eqref{eqn:theoretical-discrep-definition-covariate}, denoted
\(D(\lambda \mid \boldsymbol{X})\), using \(I_{r}\) samples
\((y_{i, r})_{i = 1}^{I_{r}} \sim \Q(Y \mid X_{r})\) drawn from an
importance distribution \(\Q(Y \mid X_{r})\) with
\begin{equation}
  D(\lambda \mid \boldsymbol{X}) =
    \frac{1}{R}
    \sum_{r = 1}^{R}
    \frac{1}{I_{r}}
    \sum_{i = 1}^{I_{r}}
    d(
      \Pd(y_{i, r} \mid \lambda, X_{r}),
      \tc(y_{i, r} \mid X_{r})
    )
    \frac {
      \text{d}\tc(y_{i, r} \mid X_{r})
    } {
      \text{d}\Q(y_{i, r} \mid X_{r})
    }.
  \label{eqn:practical-discrep-definition-covariate-importance}
\end{equation}\noindent
We select \(\Q(Y \mid X_{r})\) using information about the support
\(\mathcal{Y}\), and samples from \(\Pd(Y \mid \lambda, X_{r})\) and
\(\tc(Y \mid X_{r})\). Approximating
\(\tilde{N}(\lambda \mid \boldsymbol{X})\) is usually straightforward
via Monte Carlo, and we denote the corresponding estimate (or analytic
form, if available) by \(N(\lambda \mid \boldsymbol{X})\).

\hypertarget{benchmarking-and-other-empirical-considerations}{%
\subsection{Benchmarking and other empirical
considerations}\label{benchmarking-and-other-empirical-considerations}}

We show results for both the multi-objective approach and a
single-objective approach, which optimises only
\eqref{eqn:theoretical-discrep-definition-covariate} even in Stage 2.
Given \(\lambda^{*}\), we empirically assess faithfulness by comparing
the target distribution \(\tc(Y \mid X_{r})\) and the estimated optimal
prior predictive distribution \(\Pd(Y \mid \lambda^{*}, X_{r})\).
Replicability and uniqueness are more challenging to disentangle
empirically: without replicability we are unable to conclude whether the
multi-objective optimisation problem admits a unique solution. We will
first assess replicability by examining the stability of the components
of the loss in \eqref{eqn:loss-definition} across independent
replications of the optimisation procedure. When the loss is stable
across replicates, we will assess uniqueness by examining whether the
optimal prior predictive distribution \(\Pd(Y \mid \lambda^{*}, X_{r})\)
and prior \(\Pd(\theta \mid \lambda^{*}, \boldsymbol{X})\) are stable
across replicates; stability of both is good evidence of uniqueness.

\hypertarget{examples}{%
\section{Examples}\label{examples}}

\hypertarget{calibrating-a-cure-fraction-survival-model}{%
\subsection{Calibrating a cure fraction survival
model}\label{calibrating-a-cure-fraction-survival-model}}

Cure models \citep{amico_Cure_2018} for survival data are useful when a
cure mechanism is physically plausible \emph{a priori}, and when
individuals are followed up for long enough to be certain all censored
individuals in our data are ``cured''. Such lengthy follow ups are not
always possible, but a cure model remains plausible when a large
fraction of the censored observations occur after the last observed
event time. However, we cannot distinguish in the right tail of the
survival time distribution between censored uncured individuals and
genuinely cured individuals. We suppose here that we possess prior
knowledge on the fraction of individuals likely to be cured, and the
distribution of event times amongst the uncured, and seek to translate
this information into a prior. We consider the CDF-based CvM and AD
discrepancies in this example because the target distribution is of
mixed discrete/continuous type (due to censoring). Additionally, we
specify a model with a nontrivial correlation structure, about which we
wish to specify an informative prior, which is known to be challenging.

\hypertarget{target-survival-time-distribution-and-covariate-generation}{%
\paragraph{Target survival time distribution and covariate
generation}\label{target-survival-time-distribution-and-covariate-generation}}

Suppose that individuals are followed up for an average (but arbitrary)
of 21 units of time, with those who experience the event doing so a long
time before the end of follow up. Furthermore, suppose we believe that,
\emph{a priori}, \(5\%\) of the patients will be cured, with \(0.2\%\)
of events censored due to insufficient follow up.

Consider individuals \(n = 1, \ldots, N\) with event times \(Y_{n}\) and
censoring times \(C_{n}\), such that \(Y_{n} \in (0, C_{n}]\). A target
distribution that is consistent with our beliefs comprises a point mass
of \(0.05\) at \(C_{n}\), and a lognormal distribution with location
\(\mu^{\text{LN}} = \log(3)\) and scale
\(\sigma^{\text{LN}} = 2 \mathop{/} 3\) for \(Y_{n} < C_{n}\). This
choice of lognormal has \(99.8\%\) of its mass residing below 21, and
thus produces event times that are ``well separated'' from the censoring
time, as required by a cure fraction model. Denoting the lognormal CDF
with \(\text{LogNormal}(Y; \mu, \sigma^{2})\), we define the target CDF
\begin{equation}
  \tc(Y_{n} \mid C_{n}) = 0.95 \frac{
    \text{F}^{\text{LN}}(Y_{n}; \mu^{\text{LN}}, (\sigma^{\text{LN}})^{2})
  } {
    Z_{n}
  } +
  0.05 \mathbbm{1}_{\left\{Y_{n} = C_{n}\right\}}, \qquad
  Y_{n} \in (0, C_{n}],
  \label{eqn:surv-target-cdf-definition}
\end{equation}\noindent
where
\(Z_{n} = \text{LogNormal}(C_{n}; \mu^{\text{LN}}, \left(\sigma^{\text{LN}}\right)^{2})\)
is the required normalising constant.

We simulate data for this example with \(N = 50\) individuals, each with
\(4\) correlated covariates. When we consider the censoring time
\(C_{n}\), which also functions as a covariate, we have \(B = 5\)
covariates (we use \(B\) instead of \(C\) as in Section
\ref{methodology} for clarity). In line with our target distribution,
simulated censoring times are distributed such that
\(C_{n} \sim 20 + \text{Exp}(1)\). We sample a single correlation matrix
\(\boldsymbol{Q} \sim \text{LKJ}(5)\)
\citep{lewandowski_generating_2009} and subsequently covariates
\(\tilde{\mathbf{x}}_{n} \sim \text{MultiNormal}(\boldsymbol{0}, \boldsymbol{Q})\).
This results in marginally-standardised yet correlated covariates.

\hypertarget{model}{%
\subsubsection{Model}\label{model}}

A cure model for survival data, expressed in terms of its survival
function, is
\begin{equation}
  S(Y \mid X, \theta) = \pi + (1 - \pi) \tilde{S}(Y \mid \tilde{\boldsymbol{X}}, \tilde{\theta}),
  \label{eqn:cure-model-surv-def}
\end{equation}\noindent
where a proportion \(\pi \in (0, 1)\) of the population are \emph{cured}
and never experience the event of interest. The survival times for the
remaining \(1 - \pi\) proportion of the population are distributed
according to the \emph{uncured} survival function
\(\tilde{S}(Y \mid \tilde{\boldsymbol{X}}, \tilde{\theta})\). We use the
tilde in \(\tilde{\boldsymbol{X}}\) and \(\tilde{\theta}\) to denote
quantities specific to the uncured survival distribution, and denote
\(\theta = (\pi, \tilde{\theta})\) to align with our general notation.

Right censoring results in \(Y_{n} = C_{n}\). The censoring indicator
\(\delta_{n} = \mathbbm{1}_{\left\{Y_{n} < C_{n}\right\}}\) is 0 for
right censored events, and is 1 otherwise. We denote with
\(\tilde{\mathbf{x}}_{n}\) the \(n\)\textsuperscript{th} row of the
\(N \times (B - 1)\) covariate matrix \(\tilde{\boldsymbol{X}}\), which
we assume is column-wise standardised. We assume a Weibull regression
model for the uncured event times, with survival function
\begin{equation}
\begin{gathered}
  \tilde{S}(Y_{n} \mid \tilde{\theta}, \tilde{\mathbf{x}}_{n}, C_{n}) =
  \exp\left\{
    -Y_{n}^{\gamma}
    \exp\left\{\beta_{0} + \tilde{\mathbf{x}}_{n}\boldsymbol{\beta} \right\}
  \right\},
  \quad
  Y_{n} \in (0, C_{n}]
\end{gathered}
\label{eqn:weibull-surv-and-hazard-def}
\end{equation}
\noindent
with \(\tilde{\theta} = (\gamma, \beta_{0}, \boldsymbol{\beta})\). The
likelihood, with hazard
\(\tilde{h}(Y_{n} \mid \tilde{\theta}, \tilde{\mathbf{x}}_{n}, C_{n})\),
for the \(n\)\textsuperscript{th} individual is
\begin{equation}
  \begin{aligned}
    \pd(Y_{n} \mid \theta, \tilde{\mathbf{x}}_{n}, C_{n}) =
    &\left( 
      (1 - \pi) 
      \tilde{S}(Y_{n} \mid \tilde{\theta}, \tilde{\mathbf{x}}_{n}, C_{n})
      \tilde{h}(Y_{n} \mid \tilde{\theta}, \tilde{\mathbf{x}}_{n}, C_{n})
    \right)^{\delta_{n}} \\
    \quad & \times \left(
      \pi +
      (1 - \pi)
      \tilde{S}(Y_{n} \mid \tilde{\theta}, \tilde{\mathbf{x}}_{n}, C_{n})
    \right)^{1 - \delta_{n}}.
  \end{aligned}
  \label{eqn:weibull-likelihood-def}
\end{equation}\noindent
In the notation of Section \ref{methodology}, we have
\(Y = (Y_{n})_{n = 1}^{N}\) and
\(X = (C_{n}, \tilde{\mathbf{x}}_{n})_{n = 1}^{N}\), with \(X\)
including censoring times because the support of \(Y \mid X_{r}\)
depends on \(X_{r}\).

We will seek to identify optimal values of the hyperparamers
\(\lambda = (\alpha, \beta, \mu_{0},\allowbreak \sigma^{2}_{0}, s_{\beta}, \boldsymbol{\omega}, \allowbreak\boldsymbol{\eta}, \allowbreak a_{\pi}, b_{\pi})^{\top}\),
with \(\pi \sim \text{Beta}(a_{\pi}, b_{\pi})\),
\(\gamma \sim \text{Gamma}(\alpha, \beta)\),
\(\beta_{0} \sim \text{Normal}(\mu_{0}, \sigma_{0}^{2})\) and
\(\boldsymbol{\beta} \sim\allowbreak \text{MVSkewNormal}(\boldsymbol{0},\allowbreak \boldsymbol{S}, \boldsymbol{\eta})\),
with
\(\boldsymbol{S} = \text{diag}(s_{\beta}) \,\, \boldsymbol{\Omega} \,\, \text{diag}(s_{\beta})\)
where \(s_{\beta}\) is the prior marginal scale of
\(\boldsymbol{\beta}\) and \(\boldsymbol{\Omega}\) is parameterised by
\(\boldsymbol{\omega} = (\omega_{1}, \ldots, \omega_{6})^{\top} \in [-1, 1]^{6}\)
that uniquely determine its Cholesky factor. The skewness is necessary
to incorporate the nonlinear relationship between the hazard and the
effect of the covariates, and a covariance structure is used to account
for fact that not all the elements of \(\boldsymbol{\beta}\) can be
large simultaneously. Further details are in Supplement
\ref{additional-information-for-the-cure-fraction-survival-example}.

\hypertarget{results}{%
\subsubsection{Results}\label{results}}

\begin{figure}

{\centering \includegraphics{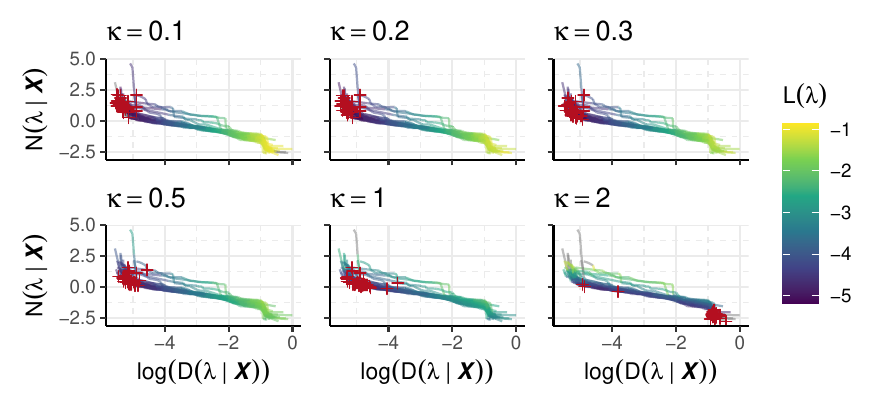} 

}

\caption{Pareto frontiers for the cure model with AD discrepancy. Each panel displays the replicates for each $\kappa$, with minimum loss point marked ($\color{myredhighlight}{+}$).}\label{fig:surv_ex_pareto_fronts}
\end{figure}

We use \(S_{r} = 10^{4}\) prior predictive samples and
\(I_{r} = 5 \times 10^3\) importance samples to estimate the
discrepancy; and optimise using \(N_{\text{CRS2}} = 2000\) CRS2
iterations, followed by \(N_{\text{batch}} = 3\) batches of Bayesian
optimisation with \(N_{\text{BO}} = 200\) iterations per batch, carrying
\(N_{\text{design}} = 60\) points between batches. We select \(\kappa\)
by inspecting the Pareto frontiers for
\(\kappa \in \{0.1, 0.2, 0.3, 0.5, 1, 2\}\) (Figure
\ref{fig:surv_ex_pareto_fronts} and Supplement
\ref{pareto-frontiers-for-cramuxe9r-von-mises-discrepancy}). Except for
the maximum and minimum values, which yield minimum loss points on the
extremes of the Pareto frontier, the minimum loss point is insensitive
to a wide range of \(\kappa\) values. We select \(\kappa = 0.3\) which
simultaneously minimises variability in loss and both objectives.

\begin{figure}

{\centering \includegraphics{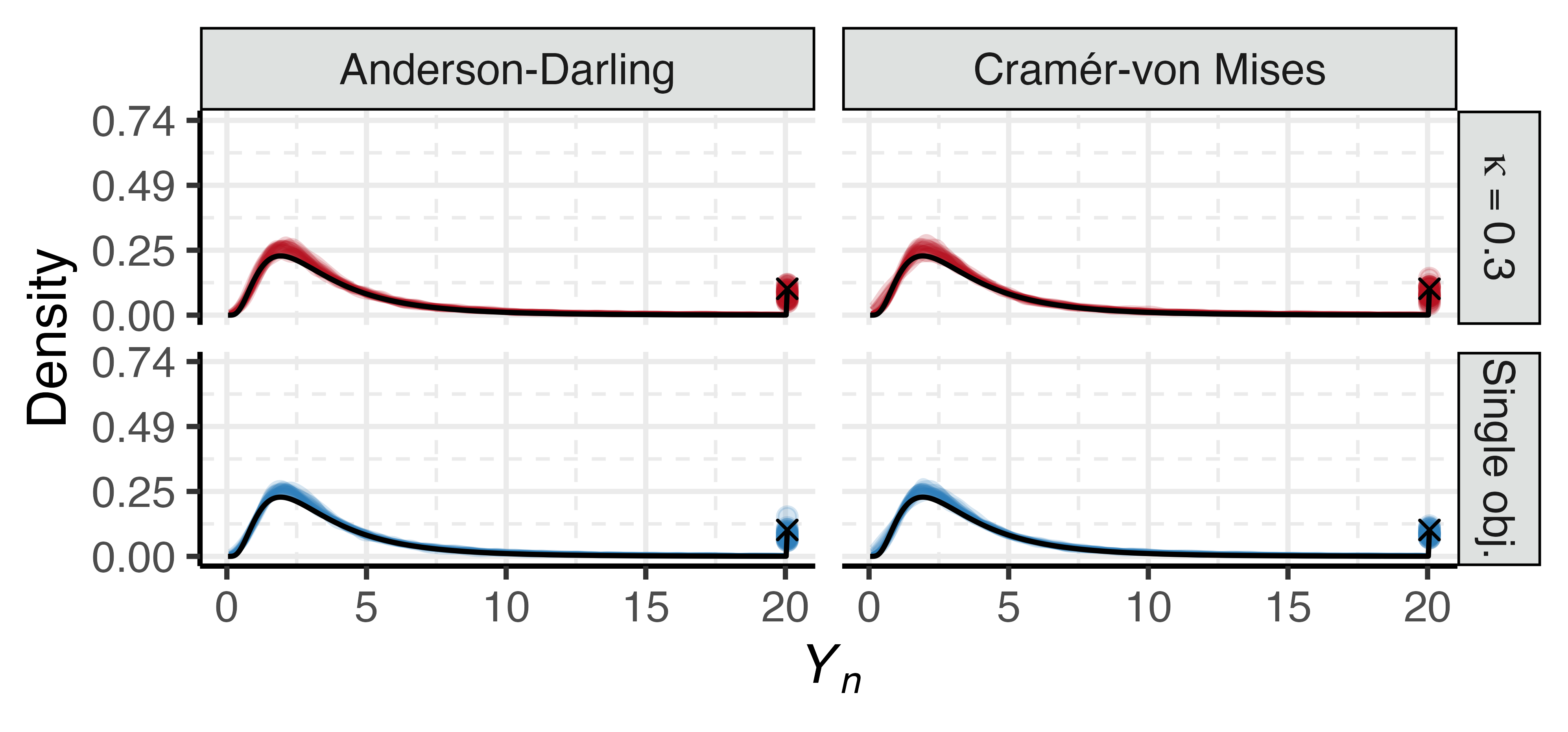} 

}

\caption{Estimated optimal prior predictive densities $\pd(Y_{n} \mid \lambda^{*})$ (\textcolor{myredhighlight}{red}/\textcolor{mymidblue}{blue} lines and dots) and target densities $\tp(Y_{n} \mid C_{N})$ (black lines and crosses).}\label{fig:surv_ex_ppd_y}
\end{figure}

The values of the loss and discrepancy functions at \(\lambda^{*}\)
across replicas are tightly distributed (see Supplement
\ref{final-objective-values}), which indicates replicability. Across all
replicates and discrepancies the estimated optimal prior predictive
distribution is highly faithful to the target, as illustrated for
individual \(n = 9\) in Figure \ref{fig:surv_ex_ppd_y} (other
individuals are visually indistinguishable).

\begin{figure}

{\centering \includegraphics{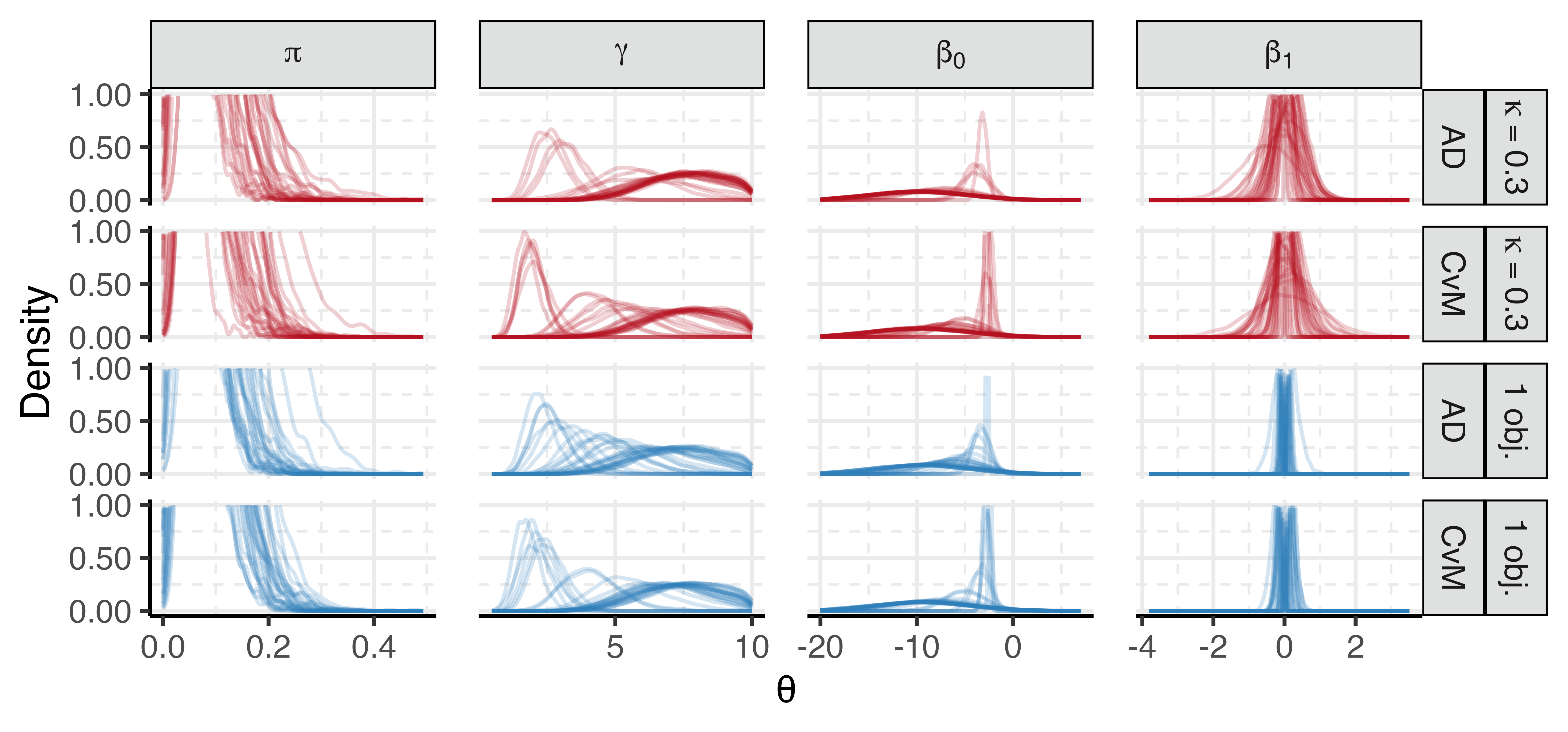} 

}

\caption{Estimated optimal prior marginal densities of $\pd(\theta \mid \lambda^{*})$ for each component of $\theta$ ($\beta_2, \dots, \beta_4$, not shown, are near identical to $\beta_1$). Both axes are truncated for readability.}\label{fig:surv_ex_ppd_theta}
\end{figure}

Figure \ref{fig:surv_ex_ppd_theta} displays the marginals of \(\theta\)
for each independent replicate. The single objective approach (i.e.~only
optimising predictive discrepancy, shown in blue) consistently locates
the degenerate, non-unique solution where all the variation in the
uncensored event times is attributed to the baseline hazard shape
\(\gamma\) and the intercept \(\beta_{0}\): i.e.~all the mass for
\(\boldsymbol{\beta}\) (the regression coefficients) is close to 0. The
combination of \(\gamma\) and \(\beta_{0}\) is far from unique, and
further calculation reveals that only the derived product
\(\gamma \exp{\beta_{0}}\) is uniquely determined. Given the
inter-replicate consistency previously observed in Supplement
\ref{final-objective-values} we infer that the solution is not unique.
In the multi-objective approach, there is a preference for the optima
surrounding \(\gamma \approx 7.5, \beta_{0} \approx -10\); an
improvement in uniqueness over the single objective approach, but
imperfect.

\begin{figure}

{\centering \includegraphics{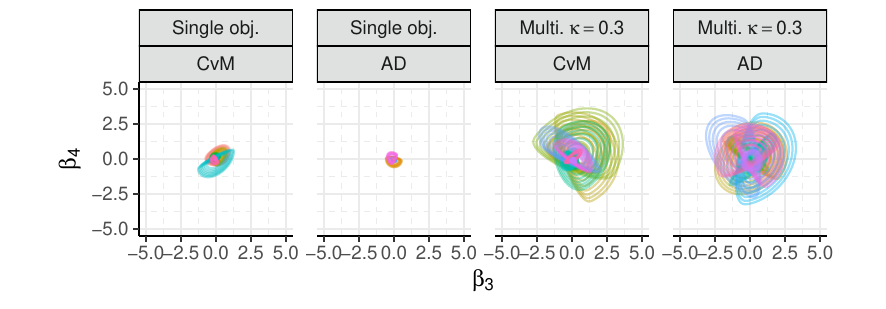} 

}

\caption{Contours of the log prior density $\log(\pd(\beta_{3}, \beta_{4} \mid \lambda^{*}))$ at the optima. For clarity we plot only the final 12 replicates, each in a unique colour.}\label{fig:surv_ex_beta_cov}
\end{figure}

Figure \ref{fig:surv_ex_beta_cov} displays the bivariate prior marginal
densities for \(\beta_{3}\) and \(\beta_{4}\), two representative
elements of \(\boldsymbol{\beta}\). Nonuniqueness is clearly apparent,
with both positive and negative marginal skewness possible. The
multi-objective approach suggests a wider distribution for
\((\beta_{3}, \beta_{4})\), as does the CvM discrepancy relative to the
AD discrepancy.

Overall, the procedure produces priors that faithfully reflect the
target, in a replicable manner. However, neither multi- or
single-objective solutions are unique, particularly for the covariance
structure, with the former closer to unique for \(\gamma\).

\hypertarget{priors-from-model-derived-quantities}{%
\subsection{Priors from model-derived
quantities}\label{priors-from-model-derived-quantities}}

Consider the linear model
\(Y = \boldsymbol{X}\boldsymbol{\beta} + \varepsilon\) for
\(n \times p\) design matrix \(\boldsymbol{X}\) and \(p\)-vector of
coefficients \(\boldsymbol{\beta}\) indexed by \(j = 1, \ldots, p\), and
where the noise \(\varepsilon\) has zero mean and variance
\(\sigma^{2}\). Suppose information about the fraction of variance
explained by the model is available -- from previous similar
experiments, or from knowledge of the measurement process -- in the form
of a plausible distribution for the coefficient of determination
\begin{equation}
  R^{2} = 1 - \frac {
    \sigma^{2}
  } {
    n^{-1} \boldsymbol{\beta}^{\top} \boldsymbol{X}^{\top} \boldsymbol{X} \boldsymbol{\beta} + \sigma^{2}
  }.
  \label{eqn:r2-definition}
\end{equation}\noindent assuming
that the columns of \(\boldsymbol{X}\) have been centred. Our aim is to
use our knowledge of \(R^{2}\) to set suitable priors for the regression
coefficients \(\beta\) conditional on \(\boldsymbol{X}\). This idea was
the inspiration for a class of shrinkage priors
\citep{zhang_variable_2018, zhang_bayesian_2022}, but we would like to
make this idea applicable to a wider selection of prior structures.

We consider three priors for the regression coefficients: a Gaussian
prior and two shrinkage priors. To demonstrate the challenge that noise
parameters pose for uniqueness we will assume
\(\varepsilon \sim N(0, \sigma^2)\) and seek to select the
hyperparameters for \(\sigma^2 \sim \text{InverseGamma}(a_{1}, b_{1})\).

The Gaussian prior
\(\beta_{j} \sim \text{N}\left(0, \frac{\sigma^{2}}{\gamma}\right)\) has
only one hyperparameter \(\gamma\), which controls the ratio of prior
variability due to \(\boldsymbol{\beta}\) to that of \(\varepsilon\).
Hence, we denote parameters
\(\boldsymbol{\theta}_{\text{GA}} = (\boldsymbol{\beta}, \sigma^{2})\),
and seek optimum values for hyperparameters
\(\boldsymbol{\lambda}_{\text{GA}} = (\gamma, a_{1}, b_{1})\).

The Dirichlet-Laplace prior (\emph{Dir. Lap.}) is defined
\citep{bhattacharya_Dirichlet_2015} for the \(j\)\textsuperscript{th}
coefficient such that
\(\beta_{j} \sim \text{Laplace}\left(0, \sigma \phi_{j} \tau \right)\),
\((\phi_{1}, \ldots, \phi_{p}) \sim \text{Dirichlet}(\alpha, \ldots, \alpha)\),
\(\tau \sim \text{Gamma}(p \alpha, 1 \mathop{/} 2)\). Smaller values of
the single hyperparameter \(\alpha\) yield more sparsity in
\(\boldsymbol{\beta}\). Thus we denote
\(\boldsymbol{\lambda}_{\text{DL}} = (\alpha, a_{1}, b_{1})\) and
\(\boldsymbol{\theta}_{\text{DL}} = (\boldsymbol{\beta}, \sigma^{2}, \phi_{1}, \ldots, \phi_{p}, \tau)\).

The regularised horseshoe prior (\emph{Reg. Horse.})
\citep{piironen_sparsity_2017} has more intermediary quantities and less
linearity, increasing its flexiblity but making finding optimal
hyperparameter values more challenging. The prior is
\begin{equation*}
  \begin{gathered}
  c^{2} \sim \text{InvGamma}\left(\frac{\nu}{2}, \frac{\nu s^{2}}{2}\right), \quad
  \omega \sim \text{Cauchy}^{+}\left(0, \frac{p_{0}}{p - p_{0}} \sqrt{\frac{\sigma^{2}}{n}}\right), \\
  \delta_{j} \sim \text{Cauchy}^{+}(0, 1), \quad
\tilde{\delta}^{2}_{j} = \frac{c^{2}\delta^{2}_{j}}{c^{2} + \omega^{2}\delta^{2}_{j}}, \quad
  \beta_{j} \sim \text{N}(0, \omega^{2}\tilde{\delta}^{2}_{j}),
  \end{gathered}
  \label{eqn:regularised-horseshoe-def}
\end{equation*}
\noindent
with \(\text{Cauchy}^{+}\) denoting a Cauchy distribution truncated to
\([0, \infty)\). Whilst the regularised horseshoe is carefully designed
to make \((p_{0}, \nu, s^{2})\) interpretable and easy to choose, here
we aim to see if we can choose
\(\boldsymbol{\lambda}_{\text{HS}} = (p_{0}, \nu, s^{2}, a_{1}, b_{1})\)
to match an informative prior for \(R^{2}\). We denote
\(\boldsymbol{\theta}_{\text{HS}} = (\boldsymbol{\beta}, \sigma^{2}, c^{2}, \omega, \delta_{1}, \ldots \delta_{p})\).

\hypertarget{evaluation-setup-and-tuning-parameters}{%
\subsubsection{Evaluation setup and tuning
parameters}\label{evaluation-setup-and-tuning-parameters}}

To assess each prior's ability to faithfully encode the information
present across a wide variety of target distribution and assess the
uniqueness and replicability of the optimisation process, we consider
sixteen different \(\text{Beta}(s_{1}, s_{2})\) distributions as our
target \(\tc(R^{2})\), with
\(\{s_{1}, s_{2}\} \in \mathcal{S} \times \mathcal{S}\) and
\(\mathcal{S}\) chosen to be four exponentially-spaced values between
and including \(1 \mathop{/} 3\) and \(3\) (i.e.~equally-spaced between
\(\log(1 \mathop{/} 3)\) and \(\log(3)\)). These values represent a
variety of potential forms of the supplied target predictive
distribution for \(R^2\).

We fix \(n = 50\) and \(p = 80\) with entries in \(\boldsymbol{X}\)
drawn from a standard Gaussian distribution, and assess replicability
using 10 independent runs for each prior and target. The support
\(\Lambda\) for the hyperparameters is defined in Supplement
\ref{hyperparameter-support-lambda-faithfulness-experiment}. We use
\(S = 10^{4}\) prior predictive samples, \(I = 5 \times 10^{3}\)
importance samples from a \(\text{Uniform}(0, 1)\), and use both
\(d^{\text{AD}}\) and \(d^{\text{CvM}}\) as discrepancy functions. We
employ \(N_{\text{CRS2}} = 1000\) iterations, and subsequently perform
both single and multi-objective Bayesian optimisation for
\(N_{\text{batch}} = 1\) batch of \(N_{\text{BO}} = 150\) iterations,
using \(N_{\text{design}} = 50\) points from the first stage. The single
objective approach illustrates that differences in flexibility between
priors also induce differences in uniqueness, and highlights issues in
choosing a prior for the additive noise parameter \(\sigma^{2}\).
Choosing \(\kappa\) is challenging in the multi-objective approach, as
its value should depend on the target, the discrepancy function, and the
prior. These dependencies result in 96 possible choices of \(\kappa\),
which is an infeasible number of choices to make in this example.
Instead we fix \(\kappa = 0.5\) for all multi-objective settings. We use
the secondary objective \eqref{eqn:second-objective-def}, except for
quantities where the standard deviation is undefined for some
\(\lambda \in \Lambda\), for which we use a robust scale estimator
\citep{rousseeuw_alternatives_1993}.

\hypertarget{results-1}{%
\subsubsection{Results}\label{results-1}}

\begin{figure}

{\centering \includegraphics{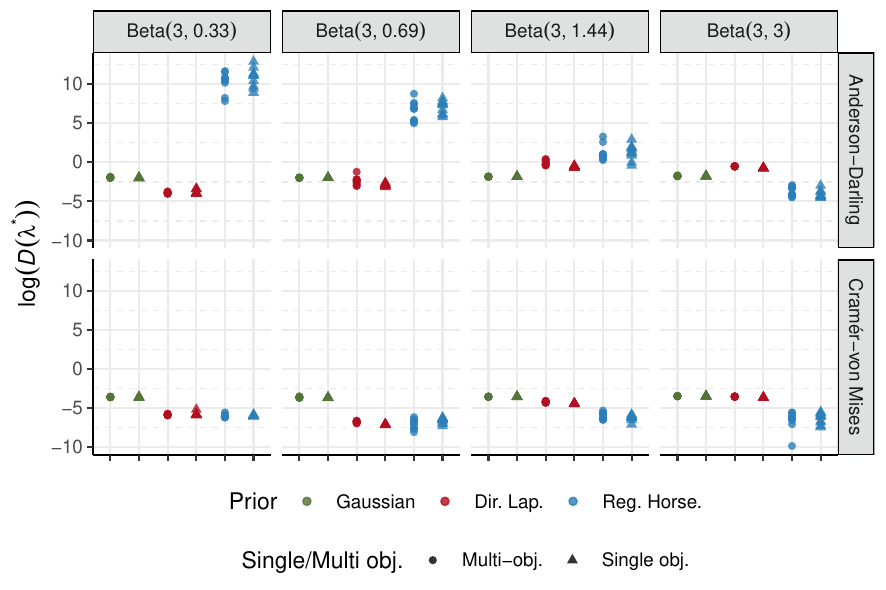} 

}

\caption{Discrepancy at the optima $\log(D(\lambda^{*}))$ for four target distributions across 10 replicates, with each the mean of 10 evaluations of $\log(D(\lambda^{*}))$ for the same $\lambda^{*}$.}\label{fig:r2_roundtrip_discrep_at_optima}
\end{figure}

We first assess replicability. It appears from Figure
\ref{fig:r2_roundtrip_discrep_at_optima} that both discrepancies are
replicable for the both Gaussian and Dirichlet-Laplace priors. In
contrast, the results for the Regularised Horseshoe prior appear to
replicate poorly under the AD discrepancy, but reasonably under the CvM
discrepancy.

We evaluate faithfulness by inspecting the densities
\(\pd(R^{2} \mid \lambda^{*})\) and \(\tp(R^{2})\) for the various
targets (all distributions in this example have corresponding
densities). A selected subset of the pairs of \((s_{1}, s_{2})\) values
are displayed in Figure \ref{fig:r2_roundtrip_full} (complete results
are in Supplement \ref{full-faithfulness-results}). The Gaussian prior
is universally poorly faithful. Both shrinkage priors perform better in
cases where one of \(s_{1}\) or \(s_{2}\) is less than 1, with the
regularised horseshoe performing better for the \(s_{1} = s_{2} > 1\)
cases. Interestingly, the results are not symmetric in \(s_{1}\) and
\(s_{2}\); the Dirichlet-Laplace prior is able to match the
\(s_{1} = 3, s_{2} = 0.69\) target well, with many of regularised
horseshoe replicates performing poorly; whilst the relative performance
is reversed for \(s_{1} = 0.69, s_{2} = 3\) (see Supplement Figure
\ref{fig:r2_roundtrip_full_supp}). There is also perceptibly more
variability in the regularised horseshoe replicates, which suggests the
optimisation problem is more challenging and the predictive discrepancy
objective is noisier. The multi-objective approach generally produces
more variable sets of optima, which is expected, as it is a more
difficult optimisation problem but we do not allow it additional
computational resources. There is little visible difference between the
CvM and AD discrepancy functions. Finally, as the values of \(s_{1}\)
and \(s_{2}\) increase, the faithfulness of the shrinkage priors
generally decreases. Across the full set of simulations, the regularised
horseshoe is evidently the most flexible.

\begin{figure}[t]

{\centering \includegraphics{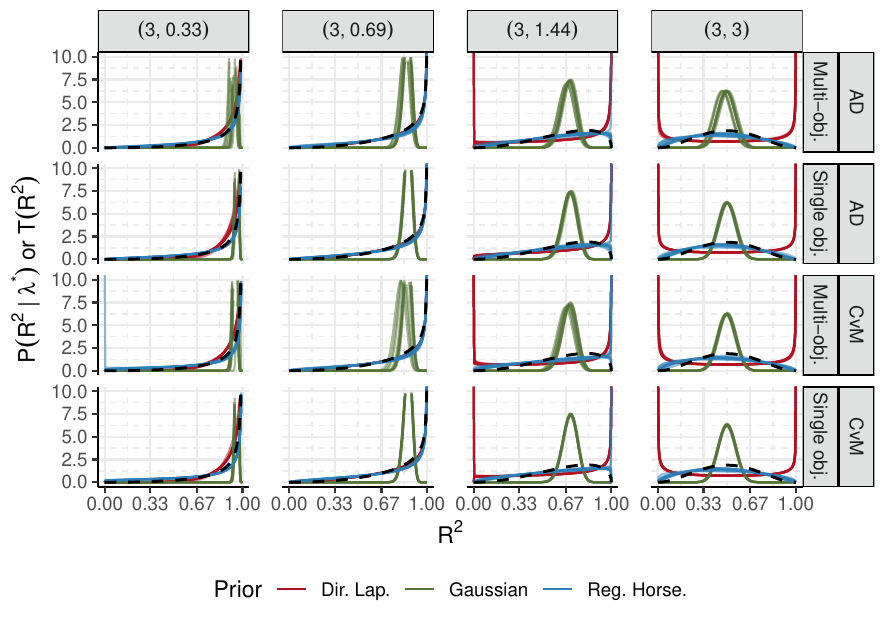} 

}

\caption{Optimal prior predictive densities $\pd(R^{2} \mid \lambda^{*})$ for the three priors considered, for selected target densities (black lines). Density values are truncated to $[0, 10]$ for readability.}\label{fig:r2_roundtrip_full}
\end{figure}

\begin{figure}

{\centering \includegraphics{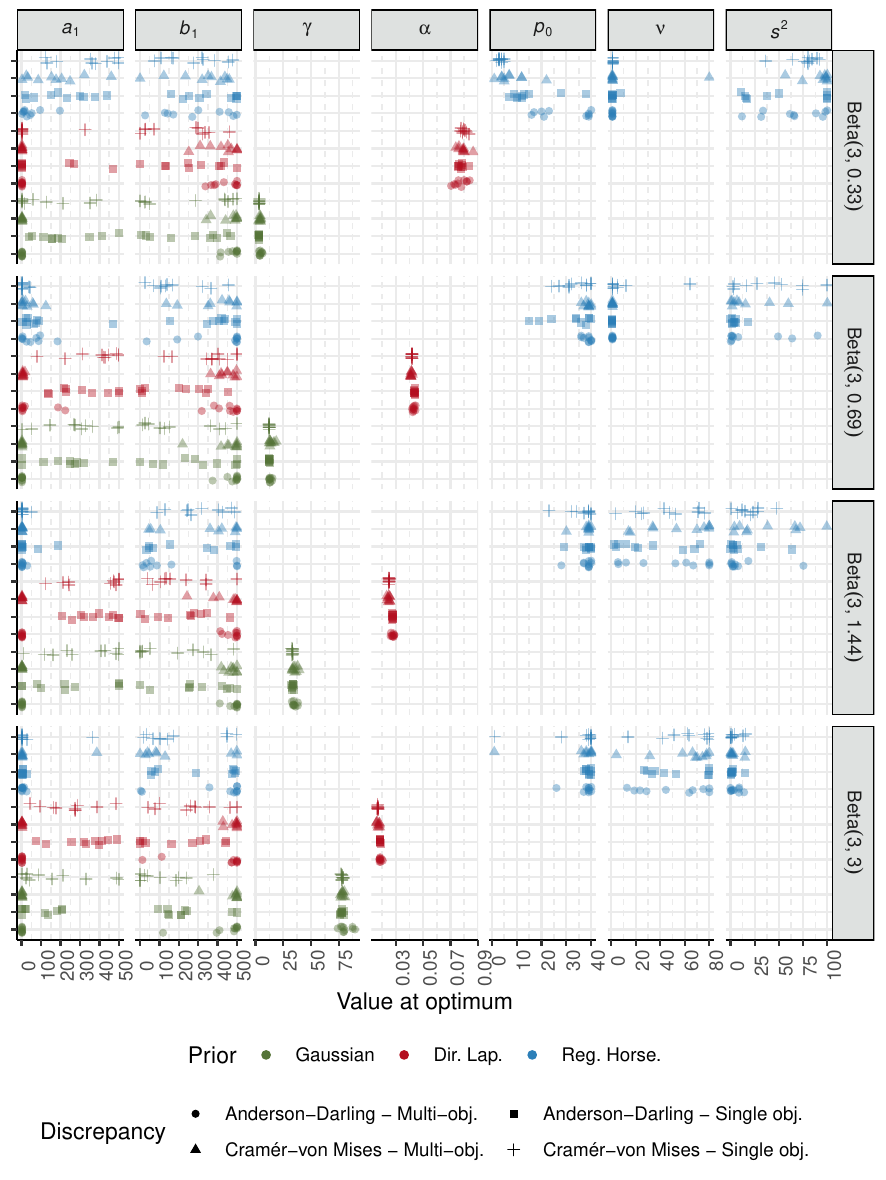} 

}

\caption{Optimal values $\lambda^{*}$ for each of the three priors considered. Columns contain (possibly prior-specific) hyperparameters, with the point colour corresponding to a specific prior. Each point's shape corresponds to the combination of discrepancy function and single or multi-objective approach. The target beta densities (denoted by the row panel titles) correspond to Figure \ref{fig:r2_roundtrip_full}.}\label{fig:r2_roundtrip_lambda}
\end{figure}

To assess uniqueness, we consider estimated optimal hyperparameter
values \(\lambda^{*}\) in each replicate. Figure
\ref{fig:r2_roundtrip_lambda} displays the estimates for \(s_{1} = 3\)
and \(s_{2} \in \{0.33, 0.69, 1.44, 3\}\), which corresponds to the
targets in Figure \ref{fig:r2_roundtrip_full}. The estimates for
\(\gamma\) and \(\alpha\), for the Gaussian and Dirichlet-Laplace priors
respectively, are consistent across replicates, which suggests the
optima may be unique. This remains true even for targets where the prior
is not faithful to the target, e.g.~the \(\text{Beta}(3, 3)\) target.
There is more variability in the hyperparameters of the regularised
horseshoe prior. There does appear to be unique solution for \(\nu\) for
the \(\text{Beta}(3, 0.33)\) and \(\text{Beta}(3, 0.69)\) targets,
whereas \(p_{0}\) and \(s^{2}\) are highly variable across replicates,
which may reflect nonuniqueness or may be due to the lack of
replicability (discussed above) of this optimisation for the regularised
horseshoe.

The hyperparameters \((a_{1}, b_{1})\) for the additive noise variance
\(\sigma^2\) are highly variable across replications for almost all
prior/target combinations. This reflects the anticipated lack of
uniqueness when incorporating such hyperparameters. It is particularly
striking for the Dirichlet-Laplace prior when
\(s_{2} \in \{0.33, 0.69\}\), where we consistently attain faithfulness
but no replicability in estimates for \((a_{1}, b_{1})\). These settings
are also interesting as the choice of single or multi-objective approach
greatly impacts the optimum values of \(a_{1}\) and \(b_{1}\).
Faithfulness of the multi-objective optima, illustrated in Figure
\ref{fig:r2_roundtrip_full}, are not appreciably worse than the single
objective approach, but the inclusion of \(\sigma^{2}\) into the
secondary objective has resulted in optimal values of \(a_{1}\) and
\(b_{1}\) that maximise the dispersion of the marginal prior (i.e.~small
\(a_{1}\) and large \(b_{1}\)). Asymptotic results are also known for
the Gaussian prior, and in Supplement
\ref{a-comparison-to-an-asymptotic-result} we further assess
replicability by benchmarking against (asymptotically) `true' values.

Our optimisation procedure has minimised \(\log(D(\lambda))\) using both
the AD and CvM discrepancy functions. The former places extra emphasis
on matching the tails of the target, and thus the Regularised Horseshoe
values in the top row of Figure \ref{fig:r2_roundtrip_discrep_at_optima}
differ from our expectations given the results in the top two rows of
Figure \ref{fig:r2_roundtrip_full}. Take, for example, the
\(s_{1} = 3, s_{2} = 0.69\) case. It is plainly evident from Figure
\ref{fig:r2_roundtrip_full} that the regularised horseshoe prior
provides a better fit to the target distribution at
\(\boldsymbol{\lambda}^{*}_{\text{HS}}\), and yet the corresponding
\(\log(D(\boldsymbol{\lambda}_{\text{HS}}^{*}))\) values in the top row
of Figure \ref{fig:r2_roundtrip_discrep_at_optima} suggest that it is
considerably worse that the Gaussian prior at
\(\boldsymbol{\lambda}_{\text{GA}}^{*}\). To reconcile this apparent
contradiction, we inspect \(\log(D(\lambda))\) at the optima computed
using the CvM discrepancy function. These values are displayed in the
bottom row of Figure \ref{fig:r2_roundtrip_discrep_at_optima}, whose
values closely match our expectations given Figure
\ref{fig:r2_roundtrip_full}. Given the range of behaviours of
\(\pd(R^{2} \mid \lambda^{*})\) for all the optima, we can conclude that
AD more heavily penalises over-estimation of the tails of
\(\pd(R^{2} \mid \lambda^{*})\) than under-estimation. This does not
discount it as an optimisation objective, but does complicate
comparisons between competing priors.

Overall, this example illustrates how information about a model-derived,
nonobservable quantity can be used to form an informative prior. The
most flexible shrinkage model (the regularised horseshoe prior) was
almost always the most faithful to the supplied information. Conversely,
the Gaussian prior is the most replicable and unique, but the lack of
faithfulness means it is unsuitable in combination with a Beta prior on
\(R^{2}\).

\hypertarget{a-human-aware-prior-for-a-human-growth-model}{%
\subsection{A human-aware prior for a human growth
model}\label{a-human-aware-prior-for-a-human-growth-model}}

Suppose an individual has their height measured at age \(t_{m}\) (in
years) for \(m = 1, \ldots, M\), with corresponding measurement
\(y_{m}\) (in centimetres). The first Preece-Baines model
\citep{preece_new_1978} for human height is the nonlinear regression
model
\begin{align}
  y_{m} &=
  h(t_{m}; \theta) + \varepsilon_{m} \label{eqn:preece-baines-model-definition-one} \\
  &= h_{1} - \frac{
    2(h_{1} - h_{0})
  } {
    \exp\{s_{0}(t_{m} - \gamma)\} +
    \exp\{s_{1}(t_{m} - \gamma)\}
  } +
  \varepsilon_{m},
  \label{eqn:preece-baines-model-definition-two}
\end{align}\noindent
with \(\varepsilon_{m} \sim \text{N}(0, \sigma^{2}_{y})\). Some
constraints are required to identify this model and ensure its physical
plausibility: specifically, we require \(0 < h_{0} < h_{1}\) and
\(0 < s_{0} < s_{1}\). To satisfy these constraints, we parameterise in
terms of \(\delta_{h} = h_{1} - h_{0}\) and
\(\delta_{s} = s_{1} - s_{0}\), which results in
\((h_{0}, \delta_{h}, s_{0}, \delta_{s})\) all sharing the same
positivity constraint. We also constrain \(\gamma\) such that
\(\gamma \in (\min_{m}(t_{m}), \max_{m}(t_{m}))\). Even with these
constraints the denominator of the fraction can be very small, yielding
negative heights, meaning the model is not plausible for all parameter
values. Furthermore, the model is poorly behaved under a flat prior, so
prior information is required to stabilise and/or regularise the
posterior.

\begin{figure}

{\centering \includegraphics{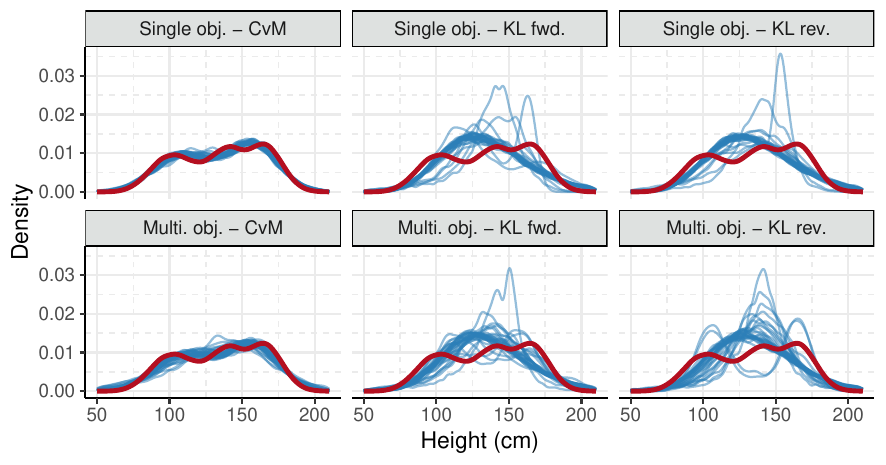} 

}

\caption{The covariate-independent marginal target density $\tp(Y)$ (\textcolor{myredhighlight}{red}) and prior predictive densities $\pd(Y \mid \lambda^{*})$ for each of the 30 replicates (\textcolor{mymidblue}{blue}).}\label{fig:pop_target_discreps}
\end{figure}

\begin{figure}

{\centering \includegraphics[width=0.8\linewidth]{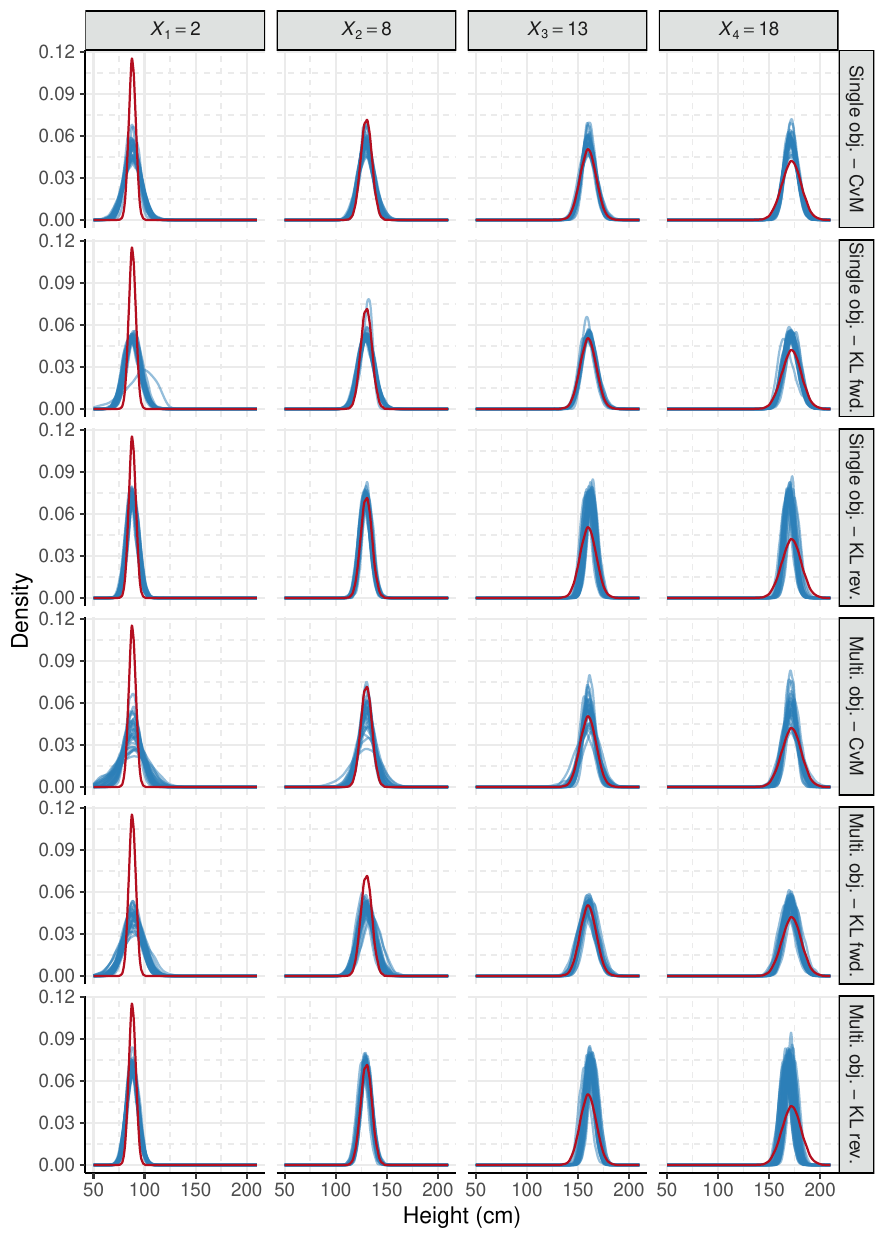} 

}

\caption{The covariate-specific target densities $\tp(Y \mid \boldsymbol{X})$ (red) and prior predictive densities $\pd(Y \mid \lambda^{*}, \boldsymbol{X})$ for each combination of discrepancy and single/multi-object approach, each of these with 30 replicates (blue lines). The columns depict the age-specific conditionals of this target.}\label{fig:cov_target_discreps}
\end{figure}

We thus seek in this example to specify priors congruent with two
specific target prior predictive distributions. We choose
\(\text{LogNormal}(\mu_{q}, s^{2}_{q})\) priors for each of the
\(q = 1, \ldots, 5\) elements of
\(\theta = (h_{0}, \delta_{h}, s_{0}, \delta_{s}, \gamma)\), and seek
optimal values of
\(\lambda = \left(\mu_{q}, s^{2}_{q}\right)_{q = 1}^{5}\) (see
Supplement \ref{hyperparameter-support-lambda-1} for \(\Lambda\)). We
fix the prior \(\sigma_{y} \sim \text{LogNormal}(0, 0.2^2)\) to avoid
uniqueness problems (Section
\ref{priors-from-model-derived-quantities}). We suppose both sex and age
(between ages 2 and 18) are uniformly distributed in our data. We first
consider a \emph{covariate-independent} prior predictive density
\(\tp(Y)\) with corresponding CDF \(\tc(Y)\) for height across the
entire age-range, derived by summarising external data. This target
(Figure \ref{fig:pop_target_discreps}) is a mixture of 3 gamma densities
specified to approximate the external data, which is multimodal due to
the fact that humans grow in spurts. We also consider a
\emph{covariate-specific} \(\tc(Y \mid X_{r})\), specifying Gaussian
height distributions at ages \(X_{r} \in (2, 8, 13, 18)\) (see Figure
\ref{fig:cov_target_discreps}, and Supplement
\ref{details-for-tcy-and-tcy-mid-x_r} for details).

\hypertarget{comparison-with-hartmann-et.-al.-and-tuning-parameters}{%
\subsubsection{Comparison with Hartmann et. al.~and tuning
parameters}\label{comparison-with-hartmann-et.-al.-and-tuning-parameters}}

\citet{hartmann_flexible_2020} also considered this example, but
elicited 6 predictive quantiles at ages \(t = (0, 2.5, 10, 17.5)\), as
opposed to entire predictive distributions at ages
\(t = (2, 8, 13, 18)\) as in our covariate-specific approach. We use
different ages because the model is stated to be accurate for ages
\(\geq 2\) \citep{preece_new_1978}. \citet{hartmann_flexible_2020} also
include in their definition of \(\theta\) a noise parameter; the
distribution of this depends on the conditional mean of the model due to
the Weibull likelihood adopted by \citet{hartmann_flexible_2020}.
Finally, \citet{hartmann_flexible_2020} elicit quantiles from 5
different users and report an estimated \(\lambda^{*}\) for each user.
These estimates (reproduced in Supplement
\ref{hartmann_flexible_2020-priors}) allow us to compare optimal the
selected priors \(\pd(\theta \mid \lambda^{*})\).

We obtain \(\lambda^{*}\) for both targets using both single- and
multi-objective optimisation processes. We use the CvM discrepancy, and
both forward and reverse KL discrepancies (numerical instability
prevented use of the AD discrepancy). We use \(S = 5 \times 10^4\)
samples from \(\pd(Y \mid \lambda)\) and likewise
\(S_{r} = 5 \times 10^4\) samples from \(\pd(Y \mid \lambda, X_{r})\)
for each of the 4 values of \(X_{r}\). We use \(I = 5 \times 10^3\) and
\(I_{r} = 5 \times 10^3\) importance samples for the CvM discrepancy,
and the same number of samples for estimating the relevant Gaussian
parameters in the KL approximation. Lastly, all settings use
\(N_{\text{CRS2}} = 2000\) CRS2 iterations, \(N_{\text{batch}} = 5\)
Bayesian optimisation batches each of \(N_{\text{BO}} = 250\)
iterations, and carry forward \(N_{\text{design}} = 50\) points per
batch. We assess replicability using 30 independent runs of each
objective/target pair.

\begin{figure}

{\centering \includegraphics{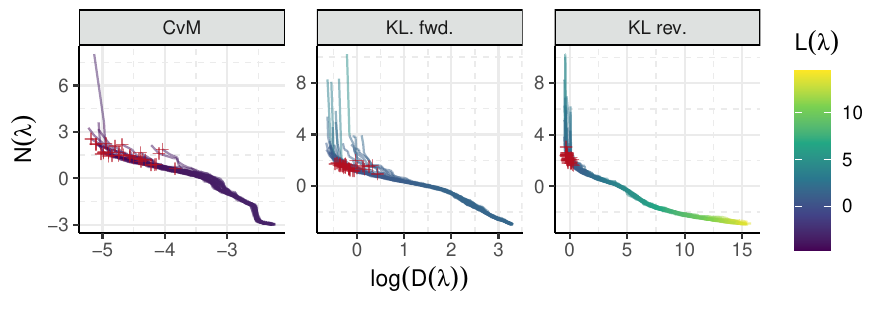} 

}

\caption{Pareto frontiers for the optimum $\kappa^{*} \in \mathcal{K}$ for each discrepancy in the \textbf{covariate-specific} example. The minimum loss point for each replicate is plotted with $\color{myredhighlight}{+}$.}\label{fig:kappa_cov}
\end{figure}

For this example, we also assess the `stability' of the resulting
posterior under each prior, by separately considering each of the 93
individuals in the \texttt{growth} data in R-package \texttt{fda}
\citep{ramsay_fda_2022}. We consider each individual's data separately,
rather than jointly, to heighten the importance of the prior. We measure
stability by whether \texttt{Stan}
\citep{stan_development_team_rstan_2021} flags a warning, setting
\texttt{adapt\_delta\ =\ 0.95} and \texttt{max\_treedepth\ =\ 12} to
minimise false positives. While a lack of warnings does not imply good
model behaviour, the presence of warning clearly indicates a problem.
This is a form of prior sensitivity analysis, but distinct from the
ideas of \citet{roos_sensitivity_2015} which consider only one
particular realisation of the data. We include the flat, improper prior
as a benchmark.

\hypertarget{results-2}{%
\subsubsection{Results}\label{results-2}}

We consider target- and discrepancy-specific ranges
\(\kappa \in \mathcal{K}\) for the multi-objective settings, and follow
our `minimum variability across replicates' heuristic (Section
\ref{algorithm-and-optimisation}) to select optimum \(\kappa^{*}\)
values (listed in Supplement \ref{choosing-kappa}). There is notable
inter-replicate variability in the Pareto frontiers at the optimal
values \(\kappa^{*}\) (Figure \ref{fig:kappa_cov}), due to the
stochasticity of our two-stage optimisation approach, with some
replicates totally dominated by other replicates. The predictive
discrepancies for the corresponding optimal \(\lambda^{*}\) values are
reasonably, but not entirely, consistent across replicates (see
Supplement
\ref{final-predictive-discrepancy-values-for-the-human-growth-example}).

Figure \ref{fig:pop_target_discreps} displays the target and prior
predictive density estimates in the covariate-independent case. The
multi-objective replicates are obtained after \(\kappa^{*}\) is chosen.
We see that introducing the secondary objective produces estimates of
\(\lambda^{*}\) that are congruent with the single objective case, but
are more variable. Both single and multi-objective approaches result in
reasonably, but not entirely, faithful densities for
\(\pd(Y \mid \lambda^{*})\), though the KL-based discrepancies are
notably less faithful. However, most optimum priors seem to accumulate
additional probability surrounding \(Y = h_{1} \approx 155\) (for the
CvM discrepancy) or \(\approx 125\) (for the KL discrepancies),
resulting in individual trajectories attaining their adult height
\(h_{1}\) for younger than expected ages \(t\) (which we will later
assess in Figure \ref{fig:regression_prior_pred}). We similarly assess
faithfulness to the covariate-specific target in Supplement
\ref{further-assessing-faithfulness}, noting that the reverse-KL
exhibits over-concentration compared to the other discrepancies (which
is to be expected, see \citet{minka_Divergence_2005}).

\begin{figure}

{\centering \includegraphics{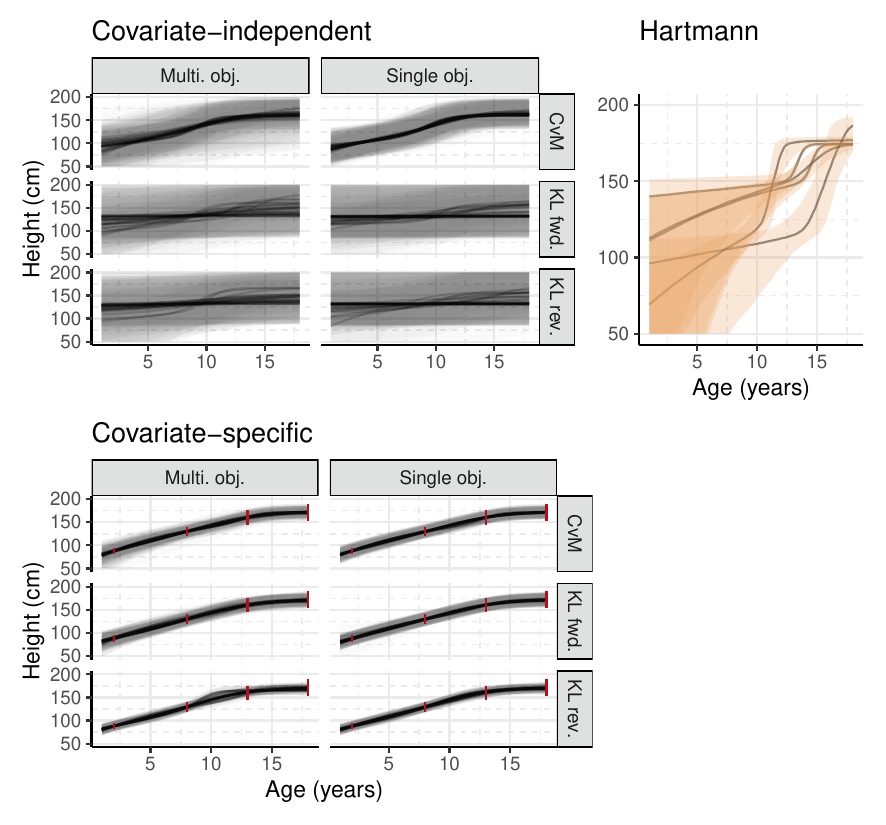} 

}

\caption{Mean (solid lines) and 95\% intervals (grey regions) for the prior mean $\pd(h(t; \theta) \mid \lambda^{*})$, for covariate-independent and covariate-specific targets in the multi- and single-objective settings, for all discrepancies; and for the Hartmann priors (with \textcolor{mywhwlow}{75\%} intervals). The y-axis is truncated to $(50, 200)$. The red lines are 95\% intervals for $\tp(Y \mid X_{r})$.}\label{fig:regression_prior_pred}
\end{figure}

Figure \ref{fig:regression_prior_pred} shows that both the
covariate-independent and covariate-specific targets yield plausible
mean growth trajectories for the CvM discrepancy, however only the
covariate-specific target does so for the KL-based discrepancies. The
covariate-independent priors are more uncertain, resulting in
implausible heights having \emph{a priori} support. The
covariate-specific priors have similar levels of uncertainty across all
ages, further suggesting that the model is too inflexible to
simultaneously match all the covariate-specific targets, which have
varying variance. All 5 of the priors from
\citet{hartmann_flexible_2020}, for a narrower uncertainty interval, are
implausible in both shape and width when viewed on this scale. It also
seems unlikely that these priors accurately reflect the information
provided by the experts in \citet{hartmann_flexible_2020}, but this
information is not reported.

\begin{figure}

{\centering \includegraphics{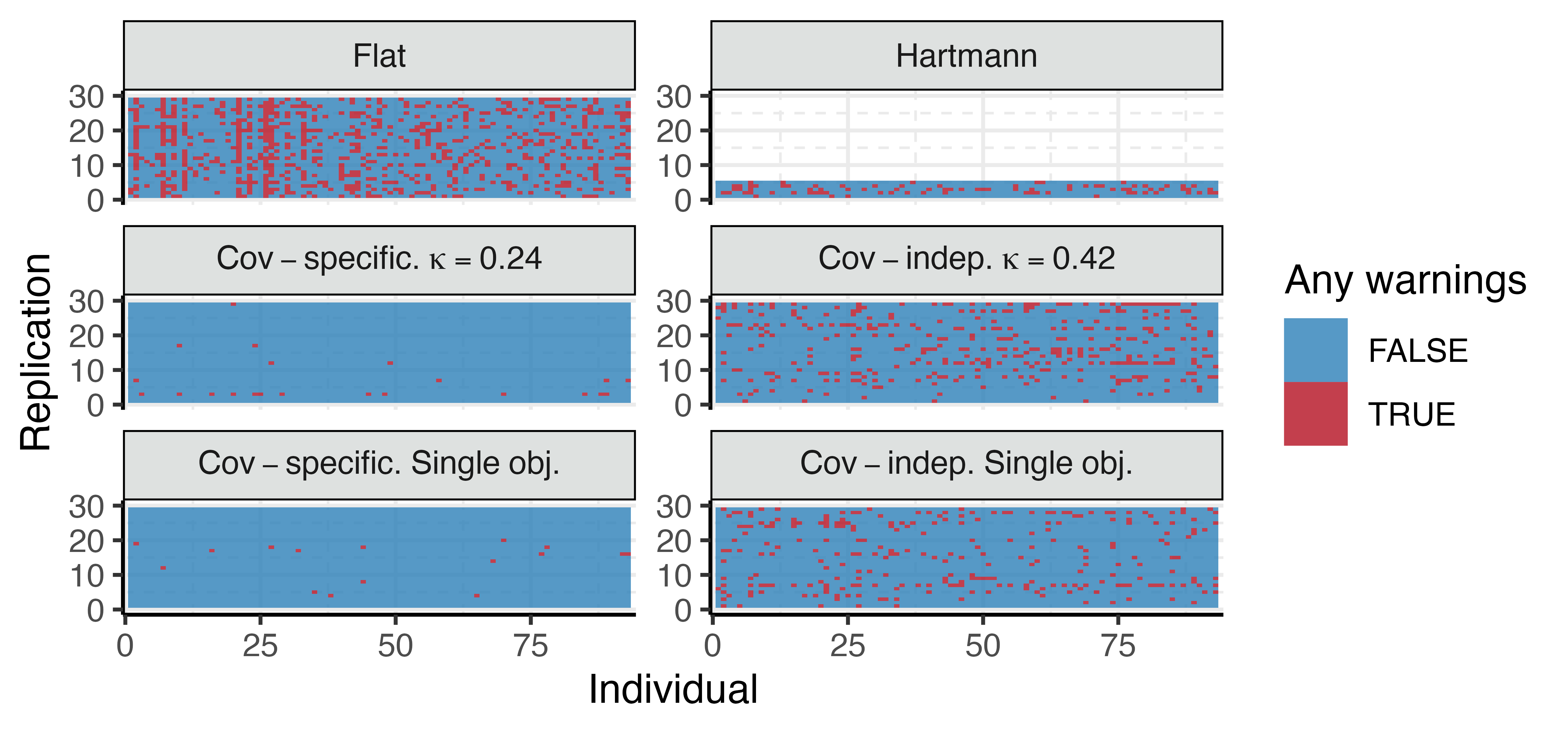} 

}

\caption{Presence/absence of \texttt{Stan} warnings for all individuals (columns) in the \texttt{growth} data and replicate prior estimates (rows). Each replicate corresponds to a run of the optimisation process (CvM discrepancy) and thus a different prior (except for flat).}\label{fig:warnings_all_priors}
\end{figure}

The different priors produce a widely ranging proportion of warning
messages in \texttt{Stan} (Figure \ref{fig:warnings_all_priors}). The
flat prior produces the most warnings, with some individuals
particularly prone to warning messages, suggesting that their data are
relatively uninformative. The Hartmann priors produce a moderate number
of warnings, with some priors less prone to produce warnings
(replications 1 and 5) than others for this dataset. Using the CvM
discrepancy, the covariate-specific approach produces fewer warnings
than the covariate-independent approach in both the single or
multi-objective cases. This reflects the additional information
available in the covariate-specific setting that results in more
informative and plausible priors. Some specific replications of the
covariate-independent approach produce many warnings, suggesting these
priors are inappropriate for many individuals.

\begin{figure}

{\centering \includegraphics{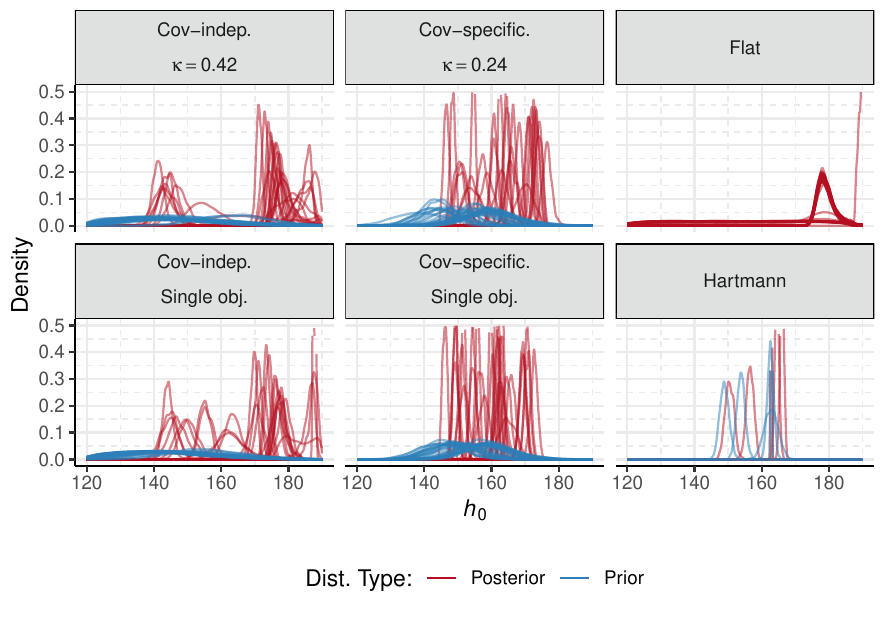} 

}

\caption{Priors (\textcolor{mymidblue}{blue}) for $h_{0}$ and corresponding posteriors (\textcolor{myredhighlight}{red}) for individual $n = 26$ using either covariate-independent and covariate-specific targets, for the CvM discrepancy; a flat prior scenario (prior not shown); and Hartmann et al. (2020).}\label{fig:small_cov_prior_post}
\end{figure}

The priors for \(h_{0}\) exhibit substantial variability across
replicates (Figure \ref{fig:small_cov_prior_post}; see Supplement
\ref{further-details-of-assessing-prior-replicability-uniqueness-and-differences-between-kl-and-cvm-discrepancies}
for a comparison to the KL discrepancy, and Supplement
\ref{full-marginal-prior-and-posterior-comparison-plots} for all
\(\theta\)). Under both covariate-independent and covariate-specific
approaches, there are two distinct unimodal priors for \(h_{0}\) with
similar loss, suggesting that \(\tc(Y \mid \boldsymbol{X})\) does not
provided enough information to uniquely determine a prior distribution.
However both priors are significantly broader than the Hartmann
et.~al.~priors. Figure \ref{fig:small_cov_prior_post} also shows the
posteriors for these parameters when using the (uninformative and thus
challenging) data from individual \(n = 26\). The posterior sampler, in
the flat prior setting, unreliably locates and adapts to the posterior,
resulting in the varied posterior estimates visible in Figure
\ref{fig:small_cov_prior_post}. Conversely, any single replicate from
the informative prior approaches provides enough regularization to
ensure consistent posterior sampling for that particular replicate.
Given this sensitivity to prior information, the posteriors strongly
depend on the prior distribution used, which as noted is not stable
under any method here. However, our priors produce posteriors for
\(\delta_{s}\) with almost all mass below 2 (Supplement
\ref{full-marginal-prior-and-posterior-comparison-plots}); this is
desirable, because \(\delta_{s} > 2\) corresponds to physiologically
implausible growth spurts that are unsupported by the data.

In summary, the priors estimated by our procedure in this example are
broadly faithful to the supplied information, except in the
covariate-specific case where model inflexibility prevents matching both
\(t = 2\) and \(t = 18\) targets simultaneously, and in the
covariate-independent case when either KL discrepancy is used. The
covariate-specific, multi-objective method appears the most useful
prior, but is arguably over concentrated, which occasionally prevents
the model from fitting the data well, although all our priors
successfully regularise the posterior sufficiently to enable accurate
posterior sampling. Our approach does not produce a unique prior,
although the secondary objective leads to a small improvement in
uniqueness (see Supplement
\ref{full-marginal-prior-and-posterior-comparison-plots}). However, some
of this non-uniqueness may be attributable to imperfect replicability of
the optimisation.

\hypertarget{conclusion}{%
\section{Conclusion}\label{conclusion}}

Setting priors for models congruent with our knowledge is often
difficult without a method for translation such as we have proposed. The
Preece-Baines model is a typical example, in which the observable is
well understood but the model parameters are not. Similarly we
anticipate our approach will be valuable for model-derived quantities
(such as \(R^{2}\)), which are often readily reasoned about but
difficult to set priors for.

One limitation of the current work is that we only partly address
non-uniqueness, but we emphasise that our methodology remains valuable
in such settings. Specifically, our approach provides insight into
consequences of certain \(\tc(Y \mid \boldsymbol{X})\): it facilitates
discovering which components of \(\lambda\) are uniquely determined, and
consideration of whether any differences between \(\tc(Y \mid X_{r})\)
and \(\Pd(Y \mid \lambda^{*}, X_{r})\) are attributable to model
inflexibility or an implausible target. We also have the opportunity to
re-assess whether we have information that we could employ to fix
certain components within \(\lambda\) (e.g.~the fixed prior for the
noise in the human height example). Another limitation of our current
work is that global optimisation methods lack guarantees of finding the
global optimum in finite time: results for CRS2 are largely empirical
and results for multi-objective Bayesian optimisation remain a topic of
research \citep[e.g.][]{chowdhury_Noregret_2021}. The generalisability
of our optimisation process thus requires further investigation.
Finally, the choice of secondary objective also invites future
investigation into alternatives: practitioners may have other principles
they wish to encode into the prior-setting process. Alternative
objectives that instead minimise the variation in only a subset of
\(\theta\) whilst maximising the remaining parameters, or objectives
that are functions of the joint distribution of \(\theta\), are avenues
for further research.

\hypertarget{acknowledgments-and-data-availability}{%
\section*{Acknowledgments and data
availability}\label{acknowledgments-and-data-availability}}
\addcontentsline{toc}{section}{Acknowledgments and data availability}

We thank Daniela De Angelis and Mevin Hooten for their feedback; and the
The Alan Turing Institute under the UK Engineering and Physical Sciences
Research Council (EPSRC) {[}EP/N510129/1{]} and the UK Medical Research
Council {[}MC\_UU\_00002/2, MC\_UU\_00002/20 and MC\_UU\_00040/04{]} for
support. No original data were generated; the \texttt{fda} package for
\texttt{R} contains the \texttt{growth} data. \texttt{R} code for the
examples is at \url{https://gitlab.com/andrew-manderson/pbbo-paper}. For
the purpose of open access, the author has applied a Creative Commons
Attribution (CC BY) licence to any Author Accepted Manuscript version
arising.

\renewcommand\thesection{S\arabic{section}}
\setcounter{section}{0}
\renewcommand*{\theHsection}{chX.\the\value{section}}

\hypertarget{r-package}{%
\section{R package}\label{r-package}}

We implement our methodology in an \texttt{R} package
\citep{rcoreteam_language_2023} \texttt{pbbo}
(\url{https://github.com/hhau/pbbo}). \texttt{pbbo} builds on top of
\texttt{mlrMBO} \citep{bischl_mlrmbo_2018} for multi-objective Bayesian
optimisation, \texttt{nlopt} and \texttt{nloptr}
\citep{ypma_nloptr_2022, johnson_nlopt_2014} for global optimisation
using CRS2 \citep{kaelo_variants_2006}, and other packages for internal
functionality and logging
\citep{wickham_welcome_2019, rowe_futilelogger_2016, maechler_rmpfr_2021}.
The code to reproduce the examples is available at
\url{https://gitlab.com/andrew-manderson/pbbo-paper}.

\hypertarget{further-notes-on-choosing-kappa}{%
\section{\texorpdfstring{Further notes on choosing
\(\kappa\)}{Further notes on choosing \textbackslash kappa}}\label{further-notes-on-choosing-kappa}}

Advantages of multi-objective optimisation are most immediately apparent
when the scales of our objectives differ markedly. Consider the
equivalent linearised approach, where we select \(\kappa\) \emph{before}
optimisation and directly optimise
\(\tilde{L}(\lambda \mid \boldsymbol{X})\). It is generally not possible
to know the range of the values of
\(\tilde{D}(\lambda \mid \boldsymbol{X})\) and
\(\tilde{N}(\lambda \mid \boldsymbol{X})\) before optimisation.
Selecting an appropriate \(\kappa\) without this knowledge is
prohibitively difficult, leaving only the computationally expensive
trial-and-error approach -- where we re-run the optimiser for each new
possible value of \(\kappa\) -- as a plausible strategy for choosing
\(\kappa\). In contrast, given \(\mathcal{P}\) it is computationally
trivial to recompute \(\lambda^{*}\) for many possible values of
\(\kappa\) \emph{after} optimisation (e.g.~each panel of Figure
\ref{fig:kappa_cov} in the main text is trivial to compute). We can thus
select \(\kappa\) in a problem-specific manner for practically no
additional computational cost to that of the multi-objective optimiser.
Note that the multi-objective optimisation approach is more expensive
that the linearised approach, but this additional cost is dwarfed by the
number of re-runs of the latter typically required to select \(\kappa\).

\hypertarget{algorithm-and-optimisation-details}{%
\section{Algorithm and optimisation
details}\label{algorithm-and-optimisation-details}}

Here we provide further details on the algorithm and optimisation
process used, before providing an overarching algorithm for the complete
methodology (Section \ref{summary-of-complete-methodology})

\hypertarget{crs2-as-an-initialiser-for-bayesian-optimisation}{%
\subsection{CRS2 as an initialiser for Bayesian
optimisation}\label{crs2-as-an-initialiser-for-bayesian-optimisation}}

Algorithm \ref{alg:crs2-algorithmic-description} describes our use of
CRS2 \citep{kaelo_variants_2006} to obtain a suitable design to
initialise the Bayesian multi-objective optimisation approach in step 2.

\begin{algorithm}[H]
  \caption{Using CRS2 to find an initial design for Bayesian optimisation}
    \label{alg:crs2-algorithmic-description}
  \begin{algorithmic}[1]
    \Require{Log total predictive discrepancy $\log(D(\lambda \mid \boldsymbol{X}))$ (evaluable using Algorithm \ref{alg:tpd-algorithmic-description}), number of CRS2 iterations to run $N_{\text{CRS2}}$, number of points in final design $N_{\text{design}}$, number of additional padding points to add for numerical stability $N_{\text{pad}}$, hyperparameter support $\Lambda$}
    \Statex
    \Function{Initial design}{$N_{\text{CRS2}}, N_{\text{design}}, N_{\text{pad}}$}
      \State Initialise $\mathcal{S} = \{\}$, an empty set to hold possible design points 
      \For{$i$ in $1 \ldots N_{\text{CRS2}}$}
        \State Minimising $\log(D(\lambda \mid \boldsymbol{X}))$, get the $i$\textsuperscript{th} trial point $\tilde{\lambda}_{i}$ and value $\log(D(\lambda_{i} \mid \boldsymbol{X}))$ from CRS2 with local mutation \citep{kaelo_variants_2006}
        \State Compute $\tilde{w}_{i} = - \exp\left\{\log\left(D(\tilde{\lambda}_{i} \mid \boldsymbol{X})\right)\right\}$
        \State Concatenate $\mathcal{S} = \mathcal{S} \cup \left\{\tilde{\lambda}_{i}, \log(D(\tilde{\lambda}_{i} \mid \boldsymbol{X})), \tilde{w}_{i}\right\}$
      \EndFor
      \State Normalise weights such that $w_{i} = \exp\left\{\tilde{w}_{i} - \log\left(\sum_{i = 1}^{N_{\text{CRS2}}} \exp\left\{\tilde{w}_{i}\right\}\right)\right\}$ 
      \State Subsample without replacement $N_{\text{design}}$ values from $\mathcal{S}$ according to the normalised weights, and store in $\mathcal{D} = \{\lambda_{i}, \log(D(\lambda_{i} \mid \boldsymbol{X}))\}_{i = 1}^{N_{\text{design}}}$
      \State Sample $N_{\text{pad}}$ points from a Latin hypercube design spanning $\Lambda$ \citep{stein_large_1987}, evaluate $\log(D(\lambda \mid \boldsymbol{X}))$ at these points, and add them to $\mathcal{D}$
      \State \textbf{return:} $\mathcal{D} = \{\lambda_{i}, \log(D(\lambda_{i} \mid \boldsymbol{X}))\}_{i = 1}^{N_{\text{design}} + N_{\text{pad}}}$
    \EndFunction
  \end{algorithmic}
\end{algorithm}



\noindent

\hypertarget{mspot}{%
\subsection{MSPOT}\label{mspot}}

Algorithm \ref{alg:mspot-algorithmic-description} describes, in our
notation, the MSPOT \citep{zaefferer_mspot_2012} algorithm for two
objectives. Note that within the algorithm we suppress each objective's
dependence on \(\boldsymbol{X}\) for brevity.

\begin{algorithm}[H]
  \caption{Global two-objective Bayesian optimisation using MSPOT \citep{zaefferer_mspot_2012}}
    \label{alg:mspot-algorithmic-description}
  \begin{algorithmic}[1]
    \Require{Primary objective $D(\lambda)$, secondary objective $N(\lambda)$, initial design $\mathcal{D} = \left\{\lambda_{i}, D(\lambda_{i}), N(\lambda_{i})\right\}_{i = 1}^{N_{\text{design}} + N_{\text{pad}}}$, number of iterations $N_{\text{BO}}$, number of new points to evaluate the surrogate models at $N_{\text{new}}$, number of evaluations to add to the design within an iteration $N_{\text{eval}}$, hyperparameter support $\Lambda$}
    \Statex
    \Function{Bayesian optimisation using MSPOT}{$N_{\text{BO}}$}
      \For{$i$ in $1 \ldots N_{\text{BO}}$}
        \State Form Gaussian process (GP) approximations to $D(\lambda)$ and $N(\lambda)$ using $\mathcal{D}$
        \State Generate a new Latin hypercube design $\mathcal{N}$ of size $N_{\text{new}}$ covering $\Lambda$, such that $N_{\text{new}} \gg N_{\text{design}}$
        \For{$k$ in $1 \ldots N_{\text{new}}$}
          \State Use the GPs to estimate $\hat{D}(\lambda_{k})$ and $\hat{N}(\lambda_{k})$
          \State Add these to $\mathcal{N}$ so that $\mathcal{N}_{k} = \left\{\lambda_{k}, \hat{D}(\lambda_{k}), \hat{N}(\lambda_{k})\right\}$
        \EndFor
        \State Truncate $\mathcal{N}$ to $N_{\text{eval}}$ points according to the non-dominated sorting rank and hypervolume contribution \citep{beume_sms-emoa_2007, deb_multi-objective_2001, deb_fast_2002, beume_complexity_2009} of each point in $\left\{D(\lambda_{k}), N(\lambda_{k})\right\}_{k = 1}^{N_{\text{new}}}$ with $N_{\text{eval}} \ll N_{\text{new}}$
        \For{$j$ in $1 \ldots N_{\text{eval}}$}
          \State Evaluate the objectives $D(\lambda_{j})$ and $N(\lambda_{j})$ for $\lambda_{j} \in \mathcal{N}$
          \State Add these evaluations to $\mathcal{D} = \mathcal{D} \cup \left\{\lambda_{j}, D(\lambda_{j}), N(\lambda_{j})\right\}$
        \EndFor
      \EndFor 
      \State Compute the Pareto frontier $\mathcal{P} = \left\{\lambda_{i}, D(\lambda_{i}), N(\lambda_{i})\right\}_{i = 1}^{\lvert \mathcal{P} \rvert}$ from $\mathcal{D} = \left\{\lambda_{i}, D(\lambda_{i}), N(\lambda_{i})\right\}_{i = 1}^{N_{\text{design}} + N_{\text{pad}} + N_{\text{BO}} N_{\text{eval}}}$ \citep[see]{kung_finding_1975}
      \State \textbf{return:} $\mathcal{P}$ and $\mathcal{D}$
    \EndFunction
  \end{algorithmic}
\end{algorithm}



\hypertarget{inter-batch-resampling}{%
\subsection{Inter batch resampling}\label{inter-batch-resampling}}

Algorithm \ref{alg:resample-batch-algorithmic-description} describes our
inter-batch resampling algorithm that we occasionally adopt in stage two
of our optimisation process.

\begin{algorithm}[H]
  \caption{Resample the outputs from a previous batch to obtain a design for the current one.}
    \label{alg:resample-batch-algorithmic-description}
  \begin{algorithmic}[1]
    \Require{Pareto frontier $\mathcal{P} = \left\{\lambda_{i}, \log(D(\lambda_{i} \mid \boldsymbol{X})), N(\lambda_{i} \mid \boldsymbol{X})\right\}_{i = 1}^{\lvert \mathcal{P} \rvert}$ and all evaluated points $\mathcal{E} = \left\{\lambda_{i}, \log(D(\lambda_{i} \mid \boldsymbol{X})), N(\lambda_{i} \mid \boldsymbol{X})\right\}_{i = 1}^{\lvert \mathcal{E} \rvert}$ from previous batch (with $\lvert \mathcal{P} \rvert \ll \lvert \mathcal{E} \rvert$), number of design points $N_{\text{design}}$, number of padding points $N_{\text{pad}}$, hyperparameter support $\Lambda$}
    \Statex
    \Function{Next batch design}{$N_{\text{design}}, N_{\text{pad}}$}
      \State Initialise $\mathcal{D} = \mathcal{P}$
      \State Compute the weights $w_{i}$ for all points in $\mathcal{E}$ in the same manner as Algorithm \ref{alg:crs2-algorithmic-description} so that $\mathcal{E} = \left\{\lambda_{i}, \log(D(\lambda_{i} \mid \boldsymbol{X})), N(\lambda_{i} \mid \boldsymbol{X}), w_{i}\right\}_{i = 1}^{\lvert \mathcal{E} \rvert}$
      \State Sample without replacement $\max\left(N_{\text{design}} - \lvert \mathcal{P} \rvert, 0\right)$ points from $\mathcal{E}$ according to the weights and add these points to $\mathcal{D}$
      \State Sample $N_{\text{pad}}$ points from a Latin hypercube design covering $\Lambda$ and add these to $\mathcal{D}$
      \State \textbf{return:} $\mathcal{D}$ such that $\lvert \mathcal{D} \rvert = \max(N_{\text{design}}, \lvert \mathcal{P} \rvert) + N_{\text{pad}}$
    \EndFunction
  \end{algorithmic}
\end{algorithm}

\hypertarget{evaluating-dlambda-mid-boldsymbolx}{%
\subsection{\texorpdfstring{Evaluating
\(D(\lambda \mid \boldsymbol{X})\)}{Evaluating D(\textbackslash lambda \textbackslash mid \textbackslash boldsymbol\{X\})}}\label{evaluating-dlambda-mid-boldsymbolx}}

Algorithm \ref{alg:tpd-algorithmic-description} summarises the algorithm
used to evaluate \(D(\lambda \mid \boldsymbol{X})\), with further
explanation in the following subsections.

\begin{algorithm}[H]
  \caption{Evaluating approximate log total predictive discrepancy $\log(D(\lambda \mid \boldsymbol{X}))$}
    \label{alg:tpd-algorithmic-description}
  \begin{algorithmic}[1]
    \Require{Targets $\tc(Y \mid X_{r})$ for $r = 1, \ldots, R$; samplers for generating points from $\tc(Y \mid X_{r})$ and $\Pd(Y \mid \lambda, X_{r})$; discrepancy $d(\cdot, \cdot)$; number of samples to draw $S_{r}$; number of importance samples $I_{r}$; observable support $\mathcal{Y}$}
    \Statex
    \Function{Evaluate \textsc{log} $(D$}{$\lambda \mid \boldsymbol{X})$}
      \For{$r$ in $1 \ldots R$}
        \State Sample prior predictive $\boldsymbol{y}^{(\Pd)}_{r} = (y_{s, r}^{(\Pd)})_{s = 1}^{S_{r}} \sim \Pd(Y \mid \lambda, X_{r})$
                \State Use $\boldsymbol{y}^{(\Pd)}_{r}$ to form the ECDF $\hat{\Pd}(Y \mid \lambda, X_{r}, \boldsymbol{y}^{(\Pd)}_{r})$
        \State Sample target $\boldsymbol{y}^{(\tc)}_{r} = (y_{s, r}^{(\tc)})_{s = 1}^{S_{r}} \sim \tc(Y \mid X_{r})$
        \State Choose importance distribution $\Q(Y \mid X_{r})$ via Supplement \ref{choosing-importance-distribution-q}
        \State Sample importance points $(y_{i, r})_{i = 1}^{I_{r}} \sim \Q(Y \mid X_{r})$
      \EndFor
      \State Compute $\log(D(\lambda \mid \boldsymbol{X}))$ using Equations \eqref{eqn:log-discrep-func-def} -- \eqref{eqn:computing-log-ad} in Supplement \ref{numerical-considerations}
      \State \textbf{return:} Value of $\log(D(\lambda \mid \boldsymbol{X}))$
    \EndFunction
  \end{algorithmic}
\end{algorithm}

\hypertarget{choosing-importance-distribution-q}{%
\subsubsection{\texorpdfstring{Choosing importance distribution
\(\Q\)}{Choosing importance distribution \textbackslash Q}}\label{choosing-importance-distribution-q}}

Appropriate importance distributions are crucial to obtaining an
accurate and low variance estimate of
\(D(\lambda \mid \boldsymbol{X})\). For values of \(\lambda\) far from
optimal, \(\Pd(Y \mid \lambda, \boldsymbol{X})\) can differ considerably
from \(\tc(Y \mid \boldsymbol{X})\). Given a specific \(X_{r}\) we
require an importance distribution \(\Q(Y \mid X_{r})\) that places
substantial mass in the high probability regions of both
\(\tc(Y \mid X_{r})\) and \(\Pd(Y \mid \lambda, X_{r})\), as it is in
these regions that \(d(\cdot, \cdot)\) is largest. But we cannot exert
too much effort on finding these densities as they are specific to each
value of \(\lambda\), and must be found anew for each \(\lambda\).

We use three quantities to guide our choice of \(\Q(Y \mid X_{r})\),
these being the support \(\mathcal{Y}\), the samples
\(\boldsymbol{y}_{r}^{(\Pd)} \sim \Pd(Y \mid \lambda, X_{r})\), and the
samples \(\boldsymbol{y}_{r}^{(\tc)} \sim \tc(Y \mid X_{r})\). Of
primary concern is the support. If \(\mathcal{Y} = \mathbb{R}\) then we
use a mixture of Student-\(t_{5}\) distributions; for
\(\mathcal{Y} = \mathbb{R} = (0, \infty)\) we employ a mixture of gamma
distributions; and for \(\mathcal{Y} = (0, a]\) with known \(a\), we opt
for a mixture of Beta distributions with a discrete component at
\(Y = a\). The parameters of the mixture components are estimated using
the method of moments. Specifically, denoting the empirical mean of
\(\boldsymbol{y}_{r}^{(\Pd)}\) as \(\hat{\mu}^{(\Pd)}\) and the
empirical variance by \(\hat{v}^{(\Pd)}\), with \(\hat{\mu}^{(\tc)}\)
and \(\hat{v}^{(\tc)}\) defined correspondingly for
\(\boldsymbol{y}_{r}^{(\tc)}\), Table
\ref{tab:importance-sampling-appendix-table} details our method of
moments estimators for the mixture components.

In this paper we limit ourselves to one dimensional \(\mathcal{Y}\),
where importance sampling is mostly well behaved or can be tamed using a
reasonable amount of computation. This covers many models, and with the
covariate-specific target it includes regression models. It is harder to
elicit \(\tc(Y \mid \boldsymbol{X})\) for higher dimensional data
spaces, and the difficulties with higher dimensional importance sampling
are well known.

\begin{landscape}
\begin{table}[tbp]
\centering
\begin{tabular}{@{}ccccp{3.5cm}@{}}
\toprule
$\mathcal{Y}$ & $\Q_{r}(Y)$ & Parameter estimates & Mixture weights & Notes \\ 
\midrule
$\mathbb{R}$ & $\begin{aligned} & \pi_{1}\text{Student-}t_{5}(Y; \hat{\mu}_{1}, \hat{s}_{1}) + \\ & \pi_{2}\text{Student-}t_{5}(Y; \hat{\mu}_{2}, \hat{s}_{2}) \end{aligned}$ & $\begin{aligned} \hat{\mu}_{1} = \hat{\mu}^{(\Pd)}, &\, \hat{s}_{1} =  c \sqrt{\hat{v}^{(\Pd)}} \\ \hat{\mu}_{2} = \hat{\mu}^{(\tc)}, &\, \hat{s}_{2} = c \sqrt{\hat{v}^{(\tc)}} \\ \end{aligned}$ & $\pi_{1} = \pi_{2} = 0.5$ & $c$ defaults to $1.05$ \\[2.0em]
\cellcolor{gray!6}{$(0, \infty)$} & \cellcolor{gray!6}{$\begin{aligned} & \pi_{1}\text{Gamma}(Y; \hat{\alpha}_{1}, \hat{\beta}_{1}) + \\ & \pi_{2}\text{Gamma}(Y; \hat{\alpha}_{2}, \hat{\beta}_{2}) \end{aligned}$} & \cellcolor{gray!6}{$\begin{aligned} \hat{\alpha}_{1} = \frac{(\hat{\mu}^{(\Pd)})^{2}}{\tilde{\omega}^{(\Pd)}}, &\, \hat{\beta}_{1} = \frac{\hat{\mu}^{(\Pd)}}{\tilde{\omega}^{(\Pd)}} \\ \hat{\alpha}_{2} = \frac{(\hat{\mu}^{(\tc)})^{2}}{\tilde{\omega}^{(\tc)}}, &\, \hat{\beta}_{2} = \frac{\hat{\mu}^{(\tc)}}{\tilde{\omega}^{(\tc)}} \\ \end{aligned}$} & \cellcolor{gray!6}{$\pi_{1} = \pi_{2} = 0.5$} & \cellcolor{gray!6}{$\tilde{\omega} = \min(c^{2} \hat{v}, 10^{5})$, \newline $c$ defaults to $1.05$} \\[2.5em]
$[0, a]$ & $\begin{aligned} & \frac{\pi_{1}}{a}\text{Beta}\left(\frac{Y}{a}; \hat{a}_{1}, \hat{b}_{1}\right) + \\ & \frac{\pi_{2}}{a}\text{Beta}\left(\frac{Y}{a}; \hat{a}_{2}, \hat{b}_{2}\right) + \\ &\pi_{3} \mathbbm{1}_{\{Y = a\}} \end{aligned}$ & $\begin{aligned} \hat{a}_{1} & = \hat{\mu}^{(\Pd)} \left[\frac{\hat{\mu}^{(\Pd)}}{\tilde{\omega}^{(\Pd)}}(1 - \hat{\mu}^{(\Pd)}) - 1\right] \\ \hat{b}_{1} & = \frac{(1 - \hat{\mu}^{(\Pd)})}{\hat{\mu}^{(\Pd)}} \hat{a}_{1} \\ \hat{a}_{2} & = \hat{\mu}^{(\tc)} \left[\frac{\hat{\mu}^{(\tc)}}{\tilde{\omega}^{(\tc)}}(1 - \hat{\mu}^{(\tc)}) - 1\right] \\ \hat{b}_{2} & = \frac{(1 - \hat{\mu}^{(\tc)})}{\hat{\mu}^{(\tc)}} \hat{a}_{2} \end{aligned}$ & $\begin{gathered} \pi_{1} = \pi_{2} = 0.45 \\ \pi_{3} = 0.05 \end{gathered}$ & $\tilde{\omega} = \max(c^{2} \hat{v}, 10^{-6})$, \newline $c$ defaults to $1.05$ \\ \bottomrule
\end{tabular}
\caption[Importance distributions and method of moments estimators for their constituent parametric distributions.]{Importance distributions and method of moments estimators for their constituent parametric distributions. Note that $c$ is a user-selected tuning parameter to enable the construction of wider importance distributions.}
\label{tab:importance-sampling-appendix-table}
\end{table}
\end{landscape}

\hypertarget{numerical-considerations}{%
\subsubsection{Numerical
considerations}\label{numerical-considerations}}

For both numerical stability and optimisation performance
\citep{eriksson_scalable_2021, snoek_input_2014} we evaluate
\(D(\lambda \mid \boldsymbol{X})\) on the log scale. This is because far
from optimal values of \(\lambda\) have corresponding
\(D(\lambda \mid \boldsymbol{X})\) many orders of magnitude larger than
near optimal values of \(\lambda\). Furthermore, the Gaussian process
approximation that underlies Bayesian optimisation assumes constant
variance, necessitating a log or log-like transformation.

Suppose again that we sample
\(\boldsymbol{y}_{r}^{(\Pd)} \sim \Pd(Y \mid \lambda, X_{r})\), from
which we form the ECDF
\(\hat{\Pd}(Y \mid \lambda, X_{r}, \boldsymbol{y}_{r}^{(\Pd)})\). Having
selected an appropriate importance distribution \(\Q(Y \mid X_{r})\) and
density \(\q(Y \mid X_{r})\) using Supplement
\ref{choosing-importance-distribution-q}, and sample importance points
\((y_{i, r})_{i = 1}^{I_{r}} \sim \Q(Y \mid X_{r})\), we define the
intermediary quantity \(z(y_{i, r})\) (in the case when densities for
the target and important distribution exist, to avoid notational
complexity) as
\begin{equation}
  z(y_{i, r}) = \log\left(
    d\left(
      \hat{\Pd}(y_{i, r} \mid \lambda, X_{r}, \boldsymbol{y}_{r}^{(\Pd)}),
      \tc(y_{i, r} \mid X_{r})
    \right)
  \right)
  + \log\left(
      \tp(y_{i, r} \mid X_{r})
    \right)
  - \log\left(
    \q(y_{i, r} \mid X_{r})
  \right),
  \label{eqn:log-discrep-func-def}
\end{equation}\noindent
and then rewrite
\eqref{eqn:practical-discrep-definition-covariate-importance} in the
main text to read
\begin{equation}
  \log(D(\lambda \mid \boldsymbol{X})) =
    -\log(R) +
    \log\left(\sum_{r = 1}^{R} \exp\left\{
    -\log(I_{r}) +
    \log\left(
      \sum_{i = 1}^{I_{r}} \exp\left\{
      z(y_{i, r})
    \right\} \right) \right\} \right).
  \label{eqn:importance-discrepancy-covariate-definition}
\end{equation}\noindent
All \(\log(\sum \exp\{\cdot\})\) terms are computed using the
numerically stable form \citep{blanchard_accurately_2021}.

Accurately evaluating \(\log(d(\cdot, \cdot))\) in
\eqref{eqn:log-discrep-func-def} involves managing the discrete nature
of the ECDF (that it returns exactly zero or one for some inputs), and
using specialised functions for each discrepancy to avoid issues with
floating point arithmetic. We compute
\(\log(d^{\text{CvM}}(\cdot, \cdot))\) using
\begin{equation}
  \log\left(d^{\text{CvM}}\left(
    \hat{\Pd}(y_{i, r} \mid \lambda, X_{r}, \boldsymbol{y}_{r}^{(\Pd)}), \tc(y_{i, r} \mid X_{r})
  \right)\right) =
  2 \log\left(\left\lvert \hat{\Pd}(y_{i, r} \mid \lambda, X_{r}, \boldsymbol{y}_{r}^{(\Pd)}) - \exp\{\mathcal{T}(y_{i, r} \mid X_{r})\} \right \rvert \right),
\label{eqn:computing-log-cvm}
\end{equation}\noindent
where
\(\mathcal{T}(y_{i, r} \mid X_{r}) = \log(\tc(y_{i, r} \mid X_{r}))\).
The log-CDF (LCDF) is often more numerically accurate for improbable
values of \(y_{i, r}\), and so our methodology assumes that it is this
LCDF form in which the target distribution is supplied. However, because
the ECDF can return exact zero/one values there is no way to perform
this computation on the log scale. We thus employ high precision
floating point numbers when exponentiating the LCDF values, using
\texttt{Rmpfr} \citep{maechler_rmpfr_2021}, to avoid evaluating
\(\log(0)\).

For \(\log(d^{\text{AD}}(\cdot, \cdot))\), additional care must be taken
as the denominator of \(d^{\text{AD}}\) in
\eqref{eqn:discrepancies-definitions} in the main text tends to
underflow to zero. Thus we evaluate it using
\begin{equation}
  \begin{multlined}
  \log\left(d^{\text{AD}}\left(
    \hat{\Pd}(y_{i, r} \mid \lambda, X_{r}, \boldsymbol{y}_{r}^{(\Pd)}), \tc(y_{i, r} \mid X_{r})
  \right)\right) = \\
  2 \log\left(
    \left\lvert \hat{\Pd}(y_{i, r} \mid \lambda, X_{r}, \boldsymbol{y}_{r}^{(\Pd)}) -
    \exp\{\mathcal{T}(y_{i, r} \mid X_{r})\} \right \rvert
  \right) - \mathcal{T}(y_{i, r} \mid X_{r}) - \texttt{log1mexp}(-\mathcal{T}(y_{i, r})),
  \end{multlined}
\label{eqn:computing-log-ad}
\end{equation}\noindent
where \(\texttt{log1mexp}(x) = \log(1 - \exp\{-x\})\) is implemented by
the \texttt{Rmpfr} package \citep{maechler_accurately_2012}. Such
precision is necessary for improbably large values of \(y_{i, r}\) under
\(\tc(y_{i, r} \mid X_{r}A)\), as the CDF/LCDF often rounds to 1/0
(respectively). It is not always feasible to evaluate
\eqref{eqn:computing-log-ad} with sufficient accuracy to avoid
under/over-flow issues -- it requires a high-precision implementation of
\(\mathcal{T}(y_{i, r} \mid X_{r})\) for extreme \(y_{i, r}\) and many
additional bits of precision for both \(y_{i, r}\) and the result. In
these settings we revert to \(\log(d^{\text{CvM}}(\cdot, \cdot))\).

\hypertarget{summary-of-complete-methodology}{%
\subsection{Summary of complete
methodology}\label{summary-of-complete-methodology}}

Lastly, Algorithm \ref{alg:pbbo-overall-algorithmic-description}
summarises the entire methodology we introduce in this paper.

\begin{algorithm}[H]
  \caption{Methodology to translate prior predictive information into a prior for the parameters in a complex model}
    \label{alg:pbbo-overall-algorithmic-description}
  \begin{algorithmic}[1]
    \Require{$\log(D(\lambda \mid \boldsymbol{X}))$ (evaluable using Algorithm \ref{alg:tpd-algorithmic-description}); secondary objective $N(\lambda \mid \boldsymbol{X})$; $\kappa$; number of Bayesian optimisation iterations $N_{\text{BO}}$; number of batches $N_{\text{batch}}$; number of CRS2 iterations $N_{\text{CRS2}}$; number of importance samples per-covariate $I_{r}$; number of prior predictive samples per-covariate $S_{r}$.}
    \Statex
    \Function{pbbo}{$\kappa, N_{\text{BO}}, N_{\text{batch}}$}
      \State Minimising $\log(D(\lambda \mid \boldsymbol{X}))$ alone, compute the initial design $\mathcal{D}$ using CRS2 via Algorithm \ref{alg:crs2-algorithmic-description}
      \For{$b$ in $1 \ldots N_{\text{batch}}$}
        \State Jointly minimising $\log(D(\lambda \mid \boldsymbol{X}))$ and $N(\lambda \mid \boldsymbol{X})$, compute the $b$\textsuperscript{th} Pareto Frontier $\mathcal{P}_{b}$ and complete design $\mathcal{D}_{b}$ using Algorithm \ref{alg:mspot-algorithmic-description}, initialising with design $\mathcal{D}$
        \State Update design $\mathcal{D}$ using $\mathcal{P}_{b}$ and $\mathcal{D}_{b}$ via Algorithm \ref{alg:resample-batch-algorithmic-description}
      \EndFor
      \State With final Pareto frontier $\mathcal{P}_{N_{\text{batch}}}$, compute $\lambda^{*} = \min L(\lambda) = \min\limits_{\lambda \in \mathcal{P}_{N_{\text{batch}}}}\log(D(\lambda \mid \boldsymbol{X})) + \kappa N(\lambda \mid \boldsymbol{X})$
      \State \textbf{return:} $\lambda^{*}$
    \EndFunction
  \end{algorithmic}
\end{algorithm}



\FloatBarrier

\hypertarget{using-the-kullbackleibler-divergence-as-a-discrepancy}{%
\section{Using the Kullback--Leibler divergence as a
discrepancy}\label{using-the-kullbackleibler-divergence-as-a-discrepancy}}

Our choice of discrepancy is general but arbitrary. Another possibility
is to minimise the Kullback--Leibler divergence from the prior
predictive distribution to the target

\begin{equation}
  \tilde{D}(\lambda \mid \boldsymbol{X}) = \kldiv{\tc(Y \mid \boldsymbol{X})}{\Pd(Y \mid \lambda, \boldsymbol{X})}.
\end{equation}

For discrete \(Y\) the challenge remains, as when minimising the CvM and
AD discrepancies, estimating \(\Pd(Y \mid \lambda, \boldsymbol{X})\);
when \(Y\) is continuous we instead require an estimate of
\(\pd(Y \mid \lambda, \boldsymbol{X})\); for mixed discrete-continuous
cases, a suitable KL divergence definition is less obvious.

Suppose \(\tc(Y \mid \boldsymbol{X})\) is multivariate Gaussian with
mean \(\mu_{1}\) and covariance \(\boldsymbol{\Sigma}_{1}\) and, for a
suitable range/value of \(\lambda\), the prior predictive is
well-approximated by another multivariate Gaussian \(\hat{\Pd}(Y)\) with
mean \(\hat{\mu}_{2}\) and covariance \(\hat{\boldsymbol{\Sigma}}_{2}\).
Given the assumption that we can draw samples from the prior predictive,
this approximation is always possible. In this case, the KL divergence
from \(\hat{\Pd}\) to \(\tc\) is

\begin{equation}
\kldiv{\tc(Y \mid \boldsymbol{X})}{\hat{\Pd}(Y)} =
  \frac{1}{2}
  \left[
    \log\frac{|\hat{\boldsymbol{\Sigma}}_{2}|}{|\boldsymbol{\Sigma}_{1}|} -
    d + 
    \text{tr}\{\hat{\boldsymbol{\Sigma}}_{2}^{-1}\boldsymbol{\Sigma}_{1} \} +
    (\hat{\mu}_{2} - \mu_{1})^T \hat{\boldsymbol{\Sigma}}_{2}^{-1}(\hat{\mu}_{2} - \mu_{1})
  \right].
  \label{eqn:kl-approx-fwd}
\end{equation}

For completeness, we also implement the ``reverse'' KL divergence
\(\kldiv{\hat{\Pd}(Y)}{\tc(Y \mid \boldsymbol{X})}\) (the direction
denoted in Equation \ref{eqn:kl-approx-fwd} is referred to as the
``forward'' KL divergence). We assess the differences between these
KL-based discrepancies and the CvM discrepancy in human height growth
example, where the Gaussian approximation to the prior predictive
distribution is most appropriate.

\hypertarget{ensuring-computational-equivalence-when-using-the-kl-as-a-discrepancy}{%
\subsection{Ensuring computational equivalence when using the KL as a
discrepancy}\label{ensuring-computational-equivalence-when-using-the-kl-as-a-discrepancy}}

Our KL approximation in Equation \ref{eqn:kl-approx-fwd} means we do not
need to estimate an ECDF or perform numerical integration to compute the
discrepancy. To ensure fair comparisons between this KL-based
discrepancy and the Cramér-von Mises or Anderson-Darling discrepancies,
the Gaussian approximation to the prior predictive distribution uses the
same number of samples as the CvM and AD discrepancies use in their ECDF
estimate of \(\Pd(Y \mid \lambda, \boldsymbol{X})\). For generality, we
do not require the end-user to supply \(\mu_{1}\) and
\(\boldsymbol{\Sigma}_{1}\). We instead estimate these parameters using
samples from \(\tc(Y \mid \boldsymbol{X})\), that would be employed in
locating an appropriate importance sampling distribution. This
approximation is available in our \texttt{pbbo} \texttt{R} package.

\hypertarget{additional-information-for-the-cure-fraction-survival-example}{%
\section{Additional information for the cure fraction survival
example}\label{additional-information-for-the-cure-fraction-survival-example}}

Note that the standardisation of \(\tilde{\boldsymbol{X}}\) allows us to
use only one \(s_{\beta}\) instead of one per covariate. These elements
are transformed into \(\boldsymbol{\Omega}\) using the partial
correlation method of \citet{lewandowski_generating_2009}, also employed
by the \texttt{Stan} math library
\citep{stan_development_team_stan_2022}. The \((B - 1)\)-vector
\(\boldsymbol{\eta}\) controls, but is not equal to, the marginal
skewness for each element of \(\boldsymbol{\beta}\) using the
multivariate skew-normal definition of
\citet{azzalini_multivariate_1996}, as implemented in the \texttt{sn}
package \citep{azzalini_sn_2022}.

\hypertarget{hyperparameter-support-lambda}{%
\subsection{\texorpdfstring{Hyperparameter support
\(\Lambda\)}{Hyperparameter support \textbackslash Lambda}}\label{hyperparameter-support-lambda}}

See Table \ref{tab:surv-cap-lambda-def}

\begin{table}[h]
\centering
\begin{tabular}[t]{llll}
\toprule
Hyperparameter & Lower & Upper & \# Elements\\
\midrule
\cellcolor{gray!6}{$\alpha$} & \cellcolor{gray!6}{$\epsilon$} & \cellcolor{gray!6}{20} & \cellcolor{gray!6}{1}\\
$\beta$ & $\epsilon$ & 20 & 1\\
\cellcolor{gray!6}{$\mu_{0}$} & \cellcolor{gray!6}{-10} & \cellcolor{gray!6}{10} & \cellcolor{gray!6}{1}\\
$\sigma_{0}$ & $\epsilon$ & 10 & 1\\
\cellcolor{gray!6}{$s_{\beta}$} & \cellcolor{gray!6}{$\epsilon$} & \cellcolor{gray!6}{10} & \cellcolor{gray!6}{1}\\
$\boldsymbol{\omega}$ & -1 + $\epsilon$ & 1 - $\epsilon$ & 6\\
\cellcolor{gray!6}{$\boldsymbol{\eta}$} & \cellcolor{gray!6}{-5} & \cellcolor{gray!6}{5} & \cellcolor{gray!6}{4}\\
$a_{\pi}$ & 1 & 50 & 1\\
\cellcolor{gray!6}{$b_{\pi}$} & \cellcolor{gray!6}{1} & \cellcolor{gray!6}{50} & \cellcolor{gray!6}{1}\\
\bottomrule
\end{tabular}
\caption{Hyperparameters $\lambda$ for the cure fraction model, their upper and lower limits that define $\Lambda$, and the number of elements in the hyperparameter (which is 1 for all scalar quantities). Note that $\epsilon = 10^{-4}$ is added or subtracted to the limits to avoid degenerate estimates for $\lambda$.}
\label{tab:surv-cap-lambda-def}
\end{table}

\hypertarget{pareto-frontiers-for-cramuxe9r-von-mises-discrepancy}{%
\subsection{Pareto Frontiers for Cramér-von Mises
discrepancy}\label{pareto-frontiers-for-cramuxe9r-von-mises-discrepancy}}

\begin{figure}[H]

{\centering \includegraphics{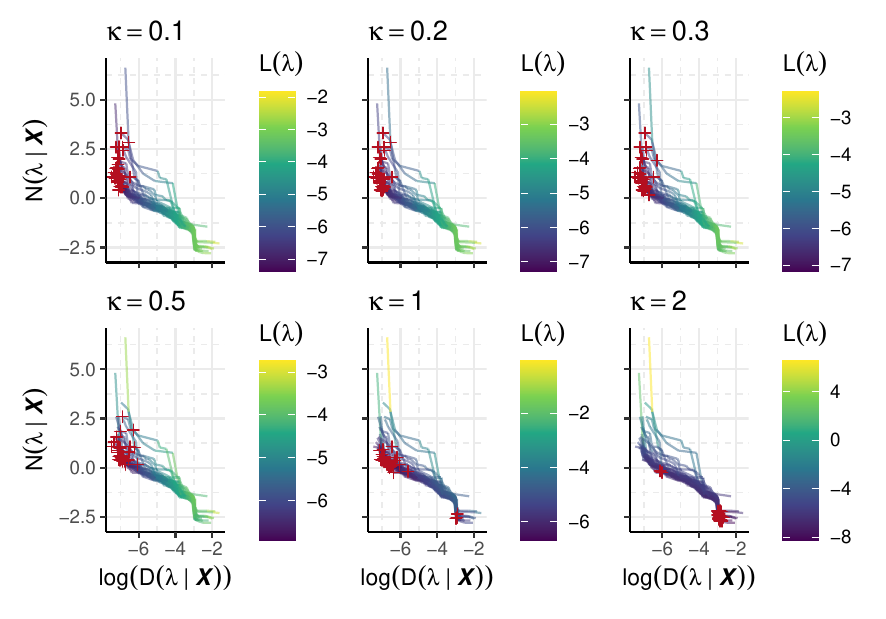} 

}

\caption{Pareto frontiers for the survival example using the Cramér-von Mises discrepancy, for the values of $\kappa$ we consider. Note that the colour scale displaying loss is panel-specific. The red crosses ($\color{myredhighlight}{+}$) indicate the minimum loss point on each frontier, for each value of $\kappa$.}\label{fig:surv_ex_pf_cvm}
\end{figure}

See Figure \ref{fig:surv_ex_pf_cvm}.

\hypertarget{final-objective-values}{%
\subsection{Final objective values}\label{final-objective-values}}

\begin{figure}[H]

{\centering \includegraphics{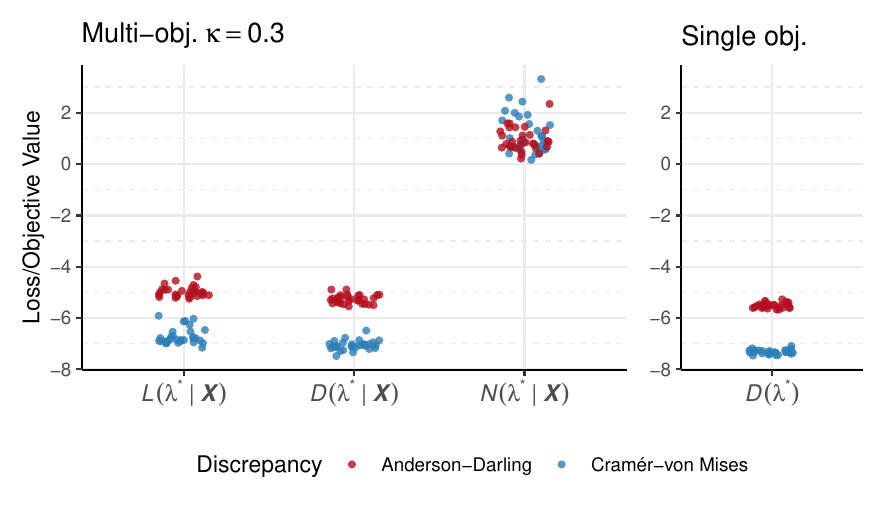} 

}

\caption{Estimates of $D(\lambda^{*} \mid \boldsymbol{X})$,  $N(\lambda^{*}\mid \boldsymbol{X})$ and $L(\lambda^{*} \mid \boldsymbol{X})$ across replicates for the cure fraction survival model.}\label{fig:surv_ex_final_objective_values}
\end{figure}

See Figure \ref{fig:surv_ex_final_objective_values}.

\FloatBarrier

\hypertarget{additional-information-for-the-r2-example}{%
\section{\texorpdfstring{Additional information for the \(R^{2}\)
example}{Additional information for the R\^{}\{2\} example}}\label{additional-information-for-the-r2-example}}

\hypertarget{hyperparameter-support-lambda-faithfulness-experiment}{%
\subsection{\texorpdfstring{Hyperparameter support \(\Lambda\) --
faithfulness
experiment}{Hyperparameter support \textbackslash Lambda -- faithfulness experiment}}\label{hyperparameter-support-lambda-faithfulness-experiment}}

See Table \ref{tab:cap-lambda-def}. Note that for the Dirichlet-Laplace
prior, \citet{zhang_variable_2018} suggest bounding
\(\alpha \in [(\max(n, p))^{-1}, 1 \mathop{/} 2]\). In our experiments
we regularly encountered optimal values of \(\alpha\) on the lower
boundary, so we use instead \(1 \mathop{/} (3\max(n, p))\) as a lower
bound.

\begin{table}[H]
\centering
\begin{tabular}[t]{llll}
\toprule
Prior & Hyperparameter & Lower & Upper\\
\midrule
\addlinespace[0.3em]
\cellcolor{gray!6}{Gaussian} & \cellcolor{gray!6}{$a_{1}$} & \cellcolor{gray!6}{2} & \cellcolor{gray!6}{500}\\
Gaussian & $b_{1}$ & 0.2 & 500\\
\cellcolor{gray!6}{Gaussian} & \cellcolor{gray!6}{$\gamma$} & \cellcolor{gray!6}{1} & \cellcolor{gray!6}{500}\\
\addlinespace[0.8em]
Dir. Lap. & $a_{1}$ & 2 & 500\\
\cellcolor{gray!6}{Dir. Lap.} & \cellcolor{gray!6}{$b_{1}$} & \cellcolor{gray!6}{0.2} & \cellcolor{gray!6}{500}\\
Dir. Lap. & $\alpha$ & $1 \mathop{/} (3 \max(n, p))$ & $1 \mathop{/} 2$ \\
\addlinespace[0.8em]
\cellcolor{gray!6}{Reg. Horse.} & \cellcolor{gray!6}{$a_{1}$} & \cellcolor{gray!6}{2} & \cellcolor{gray!6}{500}\\
Reg. Horse. & $b_{1}$ & 0.2 & 500\\
\cellcolor{gray!6}{Reg. Horse.} & \cellcolor{gray!6}{$p_{0}$} & \cellcolor{gray!6}{1} & \cellcolor{gray!6}{$p \mathop{/} 2$}\\
Reg. Horse. & $\nu$ & 1 & 80\\
\cellcolor{gray!6}{Reg. Horse.} & \cellcolor{gray!6}{$s^{2}$} & \cellcolor{gray!6}{$10^{-5}$} & \cellcolor{gray!6}{100}\\
\bottomrule
\end{tabular}
\caption{Hyperparameters $\lambda$ for the $R^{2}$ example and their upper/lower limits that define $\Lambda$.}
\label{tab:cap-lambda-def}
\end{table}

\hypertarget{full-faithfulness-results}{%
\subsection{Full faithfulness results}\label{full-faithfulness-results}}

The complete results from the faithfulness experiment are displayed in
Figure \ref{fig:r2_roundtrip_full_supp}.

\begin{figure}[H]

{\centering \includegraphics{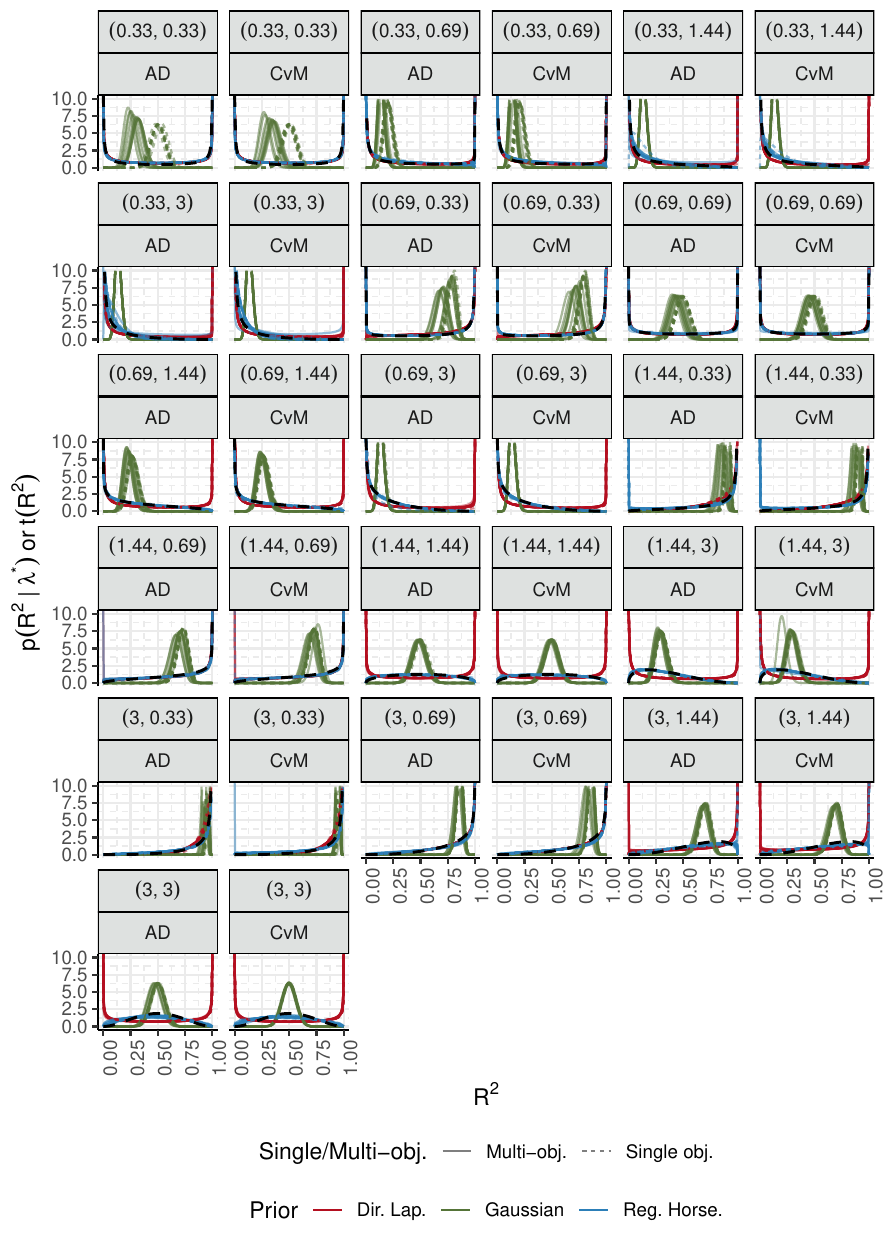} 

}

\caption{As in Figure \ref{fig:r2_roundtrip_full} (main text) but for all values of $(s_{1}, s_{2})$ denoted in the facet panels titles. The performance of the regularised horseshoe is superior to the Dirichlet-Laplace, both of which are vast improvements over the Gaussian.}\label{fig:r2_roundtrip_full_supp}
\end{figure}

\hypertarget{a-comparison-to-an-asymptotic-result}{%
\subsection{A comparison to an asymptotic
result}\label{a-comparison-to-an-asymptotic-result}}

The poor fit for the Gaussian prior observed in Figure
\ref{fig:r2_roundtrip_full} in the main text could be attributed to
issues in the optimisation process, or to the lack of flexibility in the
prior. To investigate, we compare the results for
\(\lambda_{\text{GA}}\) to Theorem 5 of \citet{zhang_variable_2018},
which is an asymptotic result regarding the optimal value of
\(\lambda_{GA}\) for a target \(\text{Beta}(s_{1}, s_{2})\) density for
\(R^{2}\). We compare pairs of \((n_{k}, p_{k})\) for
\(k = 1, \ldots, 5\), noting that assumption (A4) of Zhang and Bondell
requires that \(p_{k} = \text{o}(n_{k})\) as \(k \rightarrow \infty\)
(for strictly increasing sequences \(p_{k}\) and \(n_{k}\)). Thus we
consider values of \(p\) such that \(p_{1} = 80\) with
\(p_{k} = 2p_{k - 1}\) and \(n\) with \(n_{1} = 50\) and
\(n_{k} = n_{k - 1}^{1.2}\), both for \(k = 2, \ldots, 5\). Each
\((n_{k}, p_{k})\) pair is replicated 20 times, and for each replicate
we generate a different \(\boldsymbol{X}\) matrix with standard normal
entries. As the target density we choose \(s_{1} = 5, s_{2} = 10\) -- a
``more Gaussian'' target than previously considered and thus, we
speculate, possibly more amenable to translation with a Gaussian prior
for \(\beta\). We also use this example as an opportunity to assess if
there are notable differences between the Cramér-Von Mises discrepancy
and the Anderson-Darling discrepancy as defined in
\eqref{eqn:discrepancies-definitions} in the main text. The support
\(\Lambda\) for \(\lambda_{\text{GA}}\) differs slightly from the
example in the main text, and is defined in Table
\ref{tab:cap-lambda-def-asymp}, as matching our target with larger
design matrices requires considerably larger values of \(\gamma\).

The computation of \(R^{2}\) becomes increasingly expensive as \(n_{k}\)
and \(p_{k}\) increase, which limits the value of some of our method's
tuning parameters. The approximate discrepancy function uses
\(S = 2000\) samples from the prior predictive and is evaluated using
\(I = 500\) importance samples. We run CRS2 for
\(N_{\text{CRS2}} = 500\) iterations, using \(N_{\text{design}} = 50\)
in the initial design for the subsequent single batch of Bayesian
optimisation, which uses \(N_{\text{BO}} = 100\) iterations.

\begin{table}[H]
\centering
\begin{tabular}[t]{llll}
\toprule
Prior & Hyperparameter & Lower & Upper\\
\midrule
\addlinespace[0.3em]
Gaussian & $a_{1}$ & $2 + 10^{-6}$ & 50\\
\cellcolor{gray!6}{Gaussian} & \cellcolor{gray!6}{$b_{1}$} & \cellcolor{gray!6}{0.2} & \cellcolor{gray!6}{50}\\
Gaussian & $\gamma$ & 1 & 5000\\
\bottomrule
\end{tabular}
\caption{Hyperparameters $\lambda$ for the asymptotic example, and their upper/lower limits that define $\Lambda$.}
\label{tab:cap-lambda-def-asymp}
\end{table}

\hypertarget{results-and-analytic-comparison}{%
\subsubsection{Results and analytic
comparison}\label{results-and-analytic-comparison}}

\begin{figure}[H]

{\centering \includegraphics{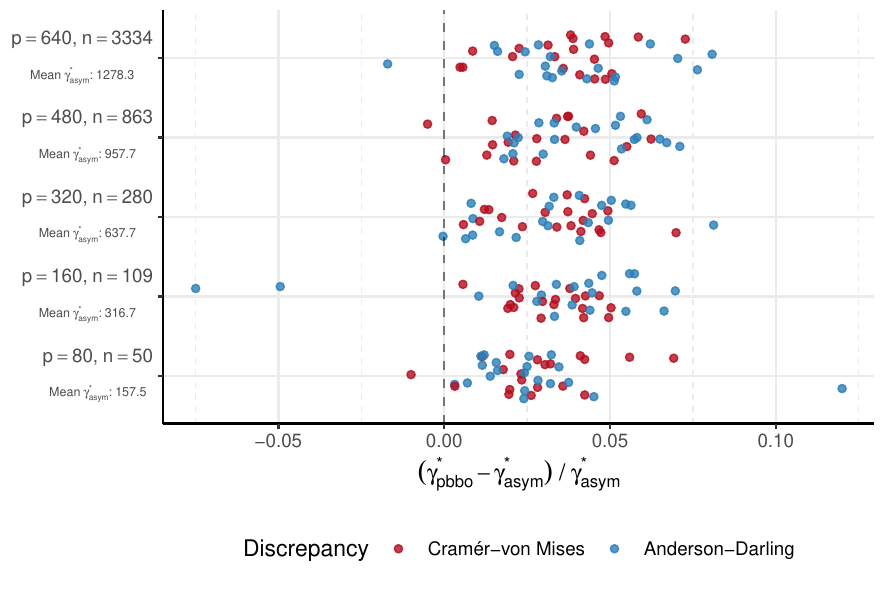} 

}

\caption{Relative difference between the value of $\gamma$ obtained using our methodology ($\gamma_{\text{pbbo}}^{*}$) and Theorem 5 of Zhang and Bondell (2018) ($\gamma_{\text{asym}}^{*}$).}\label{fig:r2_asymp_plot}
\end{figure}

Figure \ref{fig:r2_asymp_plot} displays the results in terms of the
normalised difference between the \(\gamma\) we estimate
\(\gamma_{\text{pbbo}}^{*}\), and the asymptotic result of Zhang and
Bondell \(\gamma_{\text{asym}}^{*}\). Our typical finite sample estimate
is slightly larger than the asymptotic result, and the difference
increases with \(n_{k}\) and \(p_{k}\). The variability of the
normalised difference remains roughly constant, and thus reduces on an
absolute scale, though extrema seem to occur more frequently for larger
\(n_{k}\) and \(p_{k}\). These simulations suggest that the asymptotic
regime has not been reached even at the largest \(n_{k}\) and \(p_{k}\)
values we assessed.

\begin{figure}[H]

{\centering \includegraphics{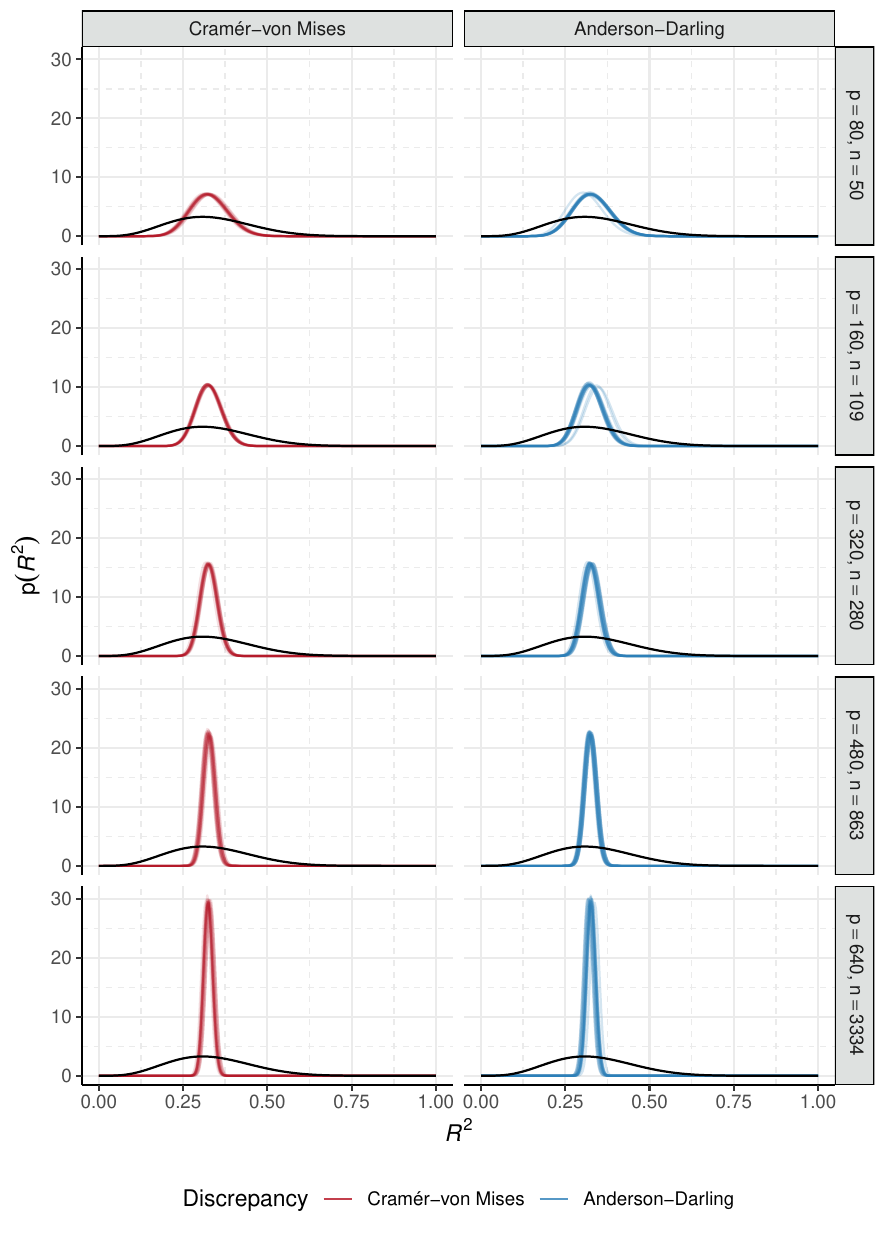} 

}

\caption{The target density $\tp(R^{2})$ and optimal prior predictive densities $\pd(R^{2} \mid \lambda^{*})$ under both the Cramér-von Mises (red, left column) and Anderson-Darling (blue, right column) discrepancies. There are 20 replicates of each discrepancy in this plot.}\label{fig:r2_target_vs_opt_prior}
\end{figure}

The estimates of \(\gamma\) are not themselves particularly
illuminating: we should instead look for differences in the distribution
of \(R^{2}\) at the optima, which is to say on the ``data'' scale.
Figure \ref{fig:r2_target_vs_opt_prior} displays the target distribution
and the prior predictive distribution at the optima
\(\pd(R^{2} \mid \lambda^{*}_{GA})\). The fit is increasingly poor as
\(n\) and \(p\) increase, and there is little difference both between
the two discrepancies and with each discrepancies replications. The lack
of difference implies that the optimisation process is consistently
locating the same minima for \(D(\lambda)\). We conclude that either 1)
the ability of the model to match the target depends on there being
additional structure in \(\boldsymbol{X}\), or 2) it is not possible to
encode the information in a \(\text{Beta}(5, 10)\) prior for \(R^{2}\)
into the Gaussian prior.

\begin{figure}[H]

{\centering \includegraphics{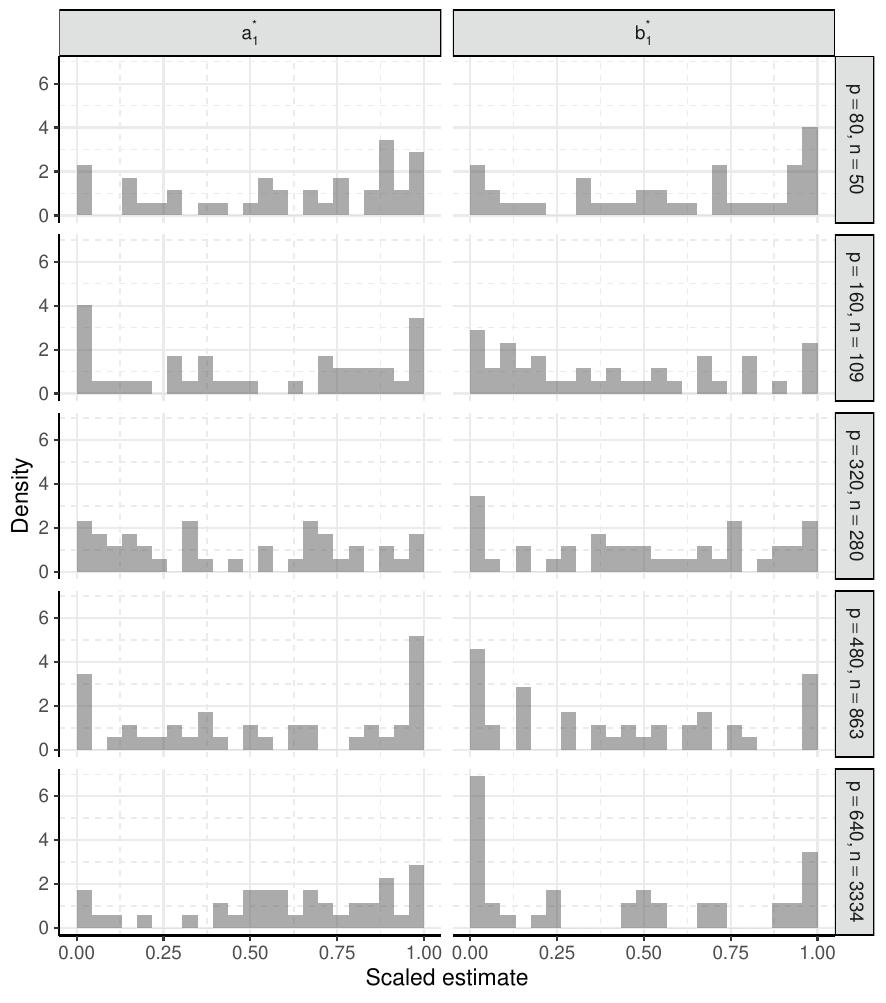} 

}

\caption{Histograms of \textit{scaled} estimates of $(a_{1}^{*}, b_{1}^{*})$ for the settings considered in Section \ref{a-comparison-to-an-asymptotic-result}. Estimates have been scaled to $[0, 1]$ for visualisation purposes using the upper and lower limits defined in Table \ref{tab:cap-lambda-def}.}\label{fig:r2_noise_hypers_plot}
\end{figure}

This example also further illustrates the difficulties inherent in
acquiring a prior for additive noise terms. Specifically, in this
example it is difficult to learn \((a_{1}, b_{1})\), despite the fact
that the contribution of \(\sigma^{2}\) in Equation
\eqref{eqn:r2-definition} in the main text is not purely additive.
However, as we see in Figure \ref{fig:r2_noise_hypers_plot}, estimates
are uniformly distributed across the permissible space, except for
bunching at the upper and lower bounds of \(\Lambda\). Note that for
numerical and computational stability, we constrain
\(a_{1} \in (2, 50]\) and \(b_{1} \in (0.2, 50]\) in this example. This
contrasts with similarity between replicates visible in Figure
\ref{fig:r2_target_vs_opt_prior}, and is thus evidence that
\((\hat{a}_{1}, \hat{b}_{1})\) have no apparent effect on the value of
\(D(\lambda^{*})\). We should instead set the prior for \(\sigma^{2}\)
based on external knowledge of the measurement process for \(Y\).

The regularisation method we employ in the two other examples in the
main text is unlikely to assist in estimating \((a_{1}, b_{1})\).
Promoting a larger mean log marginal standard deviation, with the
knowledge \(D(\lambda)\) is insensitive to the value of
\((a_{1}, b_{1})\), would simply pick the largest possible value for
\(b_{1}^{2} \mathop{/} \left((a_{1} - 1)^{2}(a_{1} - 2)\right)\), which
occurs when \(a_{1}\) is at its minimum allowable value and \(b_{1}\)
its corresponding maximum.

\FloatBarrier

\hypertarget{additional-information-for-the-preece-baines-example}{%
\section{Additional information for the Preece-Baines
example}\label{additional-information-for-the-preece-baines-example}}

\hypertarget{hyperparameter-support-lambda-1}{%
\subsection{\texorpdfstring{Hyperparameter support
\(\Lambda\)}{Hyperparameter support \textbackslash Lambda}}\label{hyperparameter-support-lambda-1}}

Table \ref{tab:pb-cap-lambda-def} contains the upper and lower limits
for each hyperparameter, thus defining the feasible region \(\Lambda\).

\begin{table}[H]
\centering
\begin{tabular}[t]{c|rrrr}
\toprule
Parameter ($\theta_{q}$) & $\mu_{q}$ -- Lower & $\mu_{q}$ -- Upper & $\sigma_{q}$ -- Lower & $\sigma_{q}$ -- Upper\\
\midrule
\cellcolor{gray!6}{$\theta_{1} = h_{0}$} & \cellcolor{gray!6}{130} & \cellcolor{gray!6}{185} & \cellcolor{gray!6}{$\epsilon$} & \cellcolor{gray!6}{30}\\
$\theta_{2} = \delta_{h}$ & $\epsilon$ & 30 & $\epsilon$ & 2\\
\cellcolor{gray!6}{$\theta_{3} = s_{0}$} & \cellcolor{gray!6}{$\epsilon$} & \cellcolor{gray!6}{0.2} & \cellcolor{gray!6}{$\epsilon$} & \cellcolor{gray!6}{0.1}\\
$\theta_{4} = \delta_{s}$ & $\epsilon$ & 1.5 & $\epsilon$ & 0.2\\
\cellcolor{gray!6}{$\theta_{5} = \gamma$} & \cellcolor{gray!6}{9} & \cellcolor{gray!6}{15} & \cellcolor{gray!6}{$\epsilon$} & \cellcolor{gray!6}{1}\\
\bottomrule
\end{tabular}
\caption{Parameter vector $\theta$ and associated model specific parameter. The rightmost four columns of the table define the upper and lower limits for the hyperparameters $(\mu_{q}, \sigma_{q})$, thus defining $\Lambda$. Informative bounds are required for numerical stability of the data generating process, and an $\epsilon = 10^{-6}$ is required to avoid nonsensical optimal values of $\lambda$.}
\label{tab:pb-cap-lambda-def}
\end{table}

\noindent

\hypertarget{details-for-tcy-and-tcy-mid-x_r}{%
\subsection{\texorpdfstring{Details for \(\tc(Y)\) and
\(\tc(Y \mid X_{r})\)}{Details for \textbackslash tc(Y) and \textbackslash tc(Y \textbackslash mid X\_\{r\})}}\label{details-for-tcy-and-tcy-mid-x_r}}

Denote with \(\text{Gamma}(Y; \alpha, \beta)\) the CDF of the gamma
distribution with shape parameter \(\alpha\) and rate \(\beta\); and
\(\text{Normal}(Y; \xi, \omega^{2})\) the CDF of the normal distribution
with mean \(\xi\) and standard deviation \(\omega\). We define the
covariate-independent target
\begin{equation}
  \tc(Y) = 0.38 \, \text{Gamma}(Y; 45.49, 0.44) +
    0.36 \, \text{Gamma}(Y; 115.41, 0.81) +
    0.27 \, \text{Gamma}(Y; 277.51, 1.64),
  \label{eqn:target-definition-pop}
\end{equation}\noindent
and the covariate-specific target
\begin{equation}
  \begin{gathered}
  \tc(Y \mid X_{1} = 2) = \text{Normal}(Y; 88, 3.5^{2}), \quad
  \tc(Y \mid X_{2} = 8) = \text{Normal}(Y; 130, 5.5^{2}), \\
  \tc(Y \mid X_{3} = 13) = \text{Normal}(Y; 160, 8^{2}), \quad
  \tc(Y \mid X_{4} = 18) = \text{Normal}(Y; 172, 9.5^{2}).
  \end{gathered}
  \label{eqn:target-definition-cov}
\end{equation}\noindent

\hypertarget{hartmann_flexible_2020-priors}{%
\subsection{\texorpdfstring{\citet{hartmann_flexible_2020}
priors}{@hartmann\_flexible\_2020 priors}}\label{hartmann_flexible_2020-priors}}

Table \ref{tab:hartmann-priors-data} contains the priors elicited by
\citet{hartmann_flexible_2020} (extracted from the supplementary
material of that paper) for the parameters in the Preece-Baines example.
To generate the prior predictive samples displayed in Figure
\ref{fig:regression_prior_pred} in the main text, we draw, for each
user, \(\theta\) from the corresponding lognormal distribution then
compute \(h(t; \theta)\) using
\eqref{eqn:preece-baines-model-definition-two} (also in the main text,
without the error term) at 250 values of \(t\) spaced evenly between
ages \(2\) and \(18\).

\begin{table}
\centering
\begin{tabular}[t]{ccrrrr}
\toprule
User & Parameter & Expectation & Variance & Lognormal $\mu$ & Lognormal $\sigma$\\
\midrule
\cellcolor{gray!6}{1} & \cellcolor{gray!6}{$h_0$} & \cellcolor{gray!6}{162.80} & \cellcolor{gray!6}{4.20} & \cellcolor{gray!6}{5.09} & \cellcolor{gray!6}{0.01}\\
1 & $h_1$ & 174.50 & 0.80 & 5.16 & 0.01\\
\cellcolor{gray!6}{1} & \cellcolor{gray!6}{$s_0$} & \cellcolor{gray!6}{0.10} & \cellcolor{gray!6}{0.10} & \cellcolor{gray!6}{-3.50} & \cellcolor{gray!6}{1.55}\\
1 & $s_1$ & 3.30 & 0.21 & 1.18 & 0.14\\
\cellcolor{gray!6}{1} & \cellcolor{gray!6}{$\theta$} & \cellcolor{gray!6}{13.40} & \cellcolor{gray!6}{0.01} & \cellcolor{gray!6}{2.60} & \cellcolor{gray!6}{0.01}\\ \addlinespace
2 & $h_0$ & 153.73 & 1.60 & 5.04 & 0.01\\
\cellcolor{gray!6}{2} & \cellcolor{gray!6}{$h_1$} & \cellcolor{gray!6}{191.74} & \cellcolor{gray!6}{4.32} & \cellcolor{gray!6}{5.26} & \cellcolor{gray!6}{0.01}\\
2 & $s_0$ & 0.04 & 0.01 & -4.21 & 1.41\\
\cellcolor{gray!6}{2} & \cellcolor{gray!6}{$s_1$} & \cellcolor{gray!6}{2.00} & \cellcolor{gray!6}{4.30} & \cellcolor{gray!6}{0.33} & \cellcolor{gray!6}{0.85}\\
2 & $\theta$ & 15.90 & 0.70 & 2.76 & 0.05\\ \addlinespace
\cellcolor{gray!6}{3} & \cellcolor{gray!6}{$h_0$} & \cellcolor{gray!6}{148.80} & \cellcolor{gray!6}{1.86} & \cellcolor{gray!6}{5.00} & \cellcolor{gray!6}{0.01}\\
3 & $h_1$ & 177.14 & 3.68 & 5.18 & 0.01\\
\cellcolor{gray!6}{3} & \cellcolor{gray!6}{$s_0$} & \cellcolor{gray!6}{0.07} & \cellcolor{gray!6}{0.00} & \cellcolor{gray!6}{-2.75} & \cellcolor{gray!6}{0.43}\\
3 & $s_1$ & 4.54 & 37.83 & 0.99 & 1.02\\
\cellcolor{gray!6}{3} & \cellcolor{gray!6}{$\theta$} & \cellcolor{gray!6}{11.31} & \cellcolor{gray!6}{0.21} & \cellcolor{gray!6}{2.42} & \cellcolor{gray!6}{0.04}\\ \addlinespace
4 & $h_0$ & 162.80 & 0.02 & 5.09 & 0.00\\
\cellcolor{gray!6}{4} & \cellcolor{gray!6}{$h_1$} & \cellcolor{gray!6}{174.50} & \cellcolor{gray!6}{0.01} & \cellcolor{gray!6}{5.16} & \cellcolor{gray!6}{0.00}\\
4 & $s_0$ & 0.10 & 0.01 & -2.65 & 0.83\\
\cellcolor{gray!6}{4} & \cellcolor{gray!6}{$s_1$} & \cellcolor{gray!6}{1.60} & \cellcolor{gray!6}{1.70} & \cellcolor{gray!6}{0.22} & \cellcolor{gray!6}{0.71}\\
4 & $\theta$ & 14.70 & 0.90 & 2.69 & 0.06\\ \addlinespace
\cellcolor{gray!6}{5} & \cellcolor{gray!6}{$h_0$} & \cellcolor{gray!6}{162.60} & \cellcolor{gray!6}{0.85} & \cellcolor{gray!6}{5.09} & \cellcolor{gray!6}{0.01}\\
5 & $h_1$ & 174.40 & 0.90 & 5.16 & 0.01\\
\cellcolor{gray!6}{5} & \cellcolor{gray!6}{$s_0$} & \cellcolor{gray!6}{0.10} & \cellcolor{gray!6}{0.01} & \cellcolor{gray!6}{-2.65} & \cellcolor{gray!6}{0.83}\\
5 & $s_1$ & 3.40 & 0.01 & 1.22 & 0.03\\
\cellcolor{gray!6}{5} & \cellcolor{gray!6}{$\theta$} & \cellcolor{gray!6}{14.60} & \cellcolor{gray!6}{0.02} & \cellcolor{gray!6}{2.68} & \cellcolor{gray!6}{0.01}\\
\bottomrule
\end{tabular}
\caption{Priors elicited by \citet{hartmann_flexible_2020} for each of the 5 users they study. \citeauthor{hartmann_flexible_2020} provide their results in the form of expected values and variances for the parameters of the model, we compute the corresponding lognormal location $\mu$ and scale $\sigma$ parameters from this information. Values are rounded to two digits of precision.}
\label{tab:hartmann-priors-data}
\end{table}

\hypertarget{choosing-kappa}{%
\subsection{\texorpdfstring{Choosing
\(\kappa^{*}\)}{Choosing \textbackslash kappa\^{}\{*\}}}\label{choosing-kappa}}

Optimal values of \(\kappa\) are selected for the multi-objective
approaches by minimising the variance of the sum of both objectives.
These values are displayed in Table \ref{tab:optimum-kappa-values}, and
are used for all multi-objective results in this section. The range
values considered, \(\mathcal{K}\), is specific to each
target/discrepancy pair, as each objective is scale-free and thus
universally applicable fixed ranges are not available.

\begin{table}

\caption{\label{tab:optimum-kappa-values}Selected optimal values $\kappa^{*}$ for each combination of target and discrepancy. Optimal values are selected as those that minimise the variance of the sum of both objectives.}
\centering
\begin{tabular}[c]{llrl}
\toprule
Target & Discrepancy & $\kappa^{*}$ & $\mathcal{K}$\\
\midrule
\cellcolor{gray!6}{Covariate-independent} & \cellcolor{gray!6}{Cramér-von Mises} & \cellcolor{gray!6}{0.42} & \cellcolor{gray!6}{$[0.05, 1.9]$}\\
Covariate-independent & KL forward & 1.43 & $[0.05, 3.5]$\\
\cellcolor{gray!6}{Covariate-independent} & \cellcolor{gray!6}{KL reverse} & \cellcolor{gray!6}{0.28} & \cellcolor{gray!6}{$[0.05, 6.0]$}\\
Covariate-dependent & Cramér-von Mises & 0.24 & $[0.05, 0.5]$\\
\cellcolor{gray!6}{Covariate-dependent} & \cellcolor{gray!6}{KL forward} & \cellcolor{gray!6}{0.49} & \cellcolor{gray!6}{$[0.05, 0.7]$}\\
Covariate-dependent & KL reverse & 0.46 & $[0.05, 2.0]$\\
\bottomrule
\end{tabular}
\end{table}

\hypertarget{pareto-frontiers-for-the-covariate-independent-target}{%
\subsection{Pareto frontiers for the covariate-independent
target}\label{pareto-frontiers-for-the-covariate-independent-target}}

\begin{figure}[H]

{\centering \includegraphics{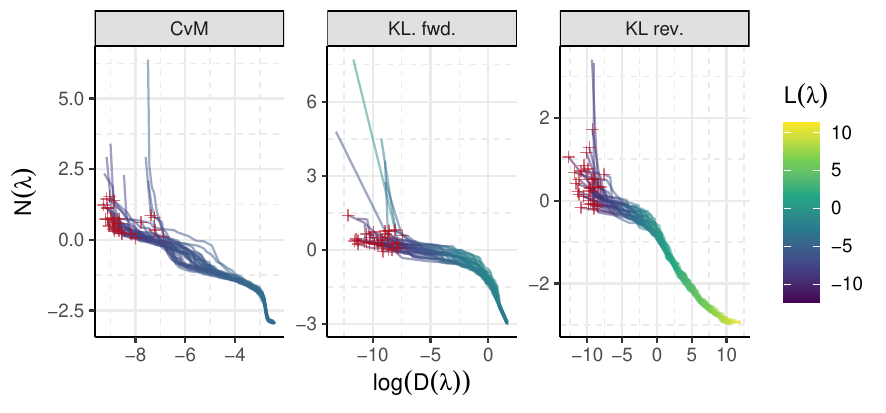} 

}

\caption{Pareto frontiers for optimum $\kappa^{*}$, listed in Table \ref{tab:optimum-kappa-values}, for the \textbf{covariate-independent} example. The minimum loss point for each replicate is plotted with $\color{myredhighlight}{+}$.}\label{fig:kappa_pop}
\end{figure}

The Pareto frontiers for the covariate-independent target for optimal
\(\kappa^{*} \in \mathcal{K}\), as defined in Table
\ref{tab:optimum-kappa-values}. is displayed in Figure
\ref{fig:kappa_pop}.

\hypertarget{final-predictive-discrepancy-values-for-the-human-growth-example}{%
\subsection{Final predictive discrepancy values for the human growth
example}\label{final-predictive-discrepancy-values-for-the-human-growth-example}}

\begin{figure}[H]

{\centering \includegraphics{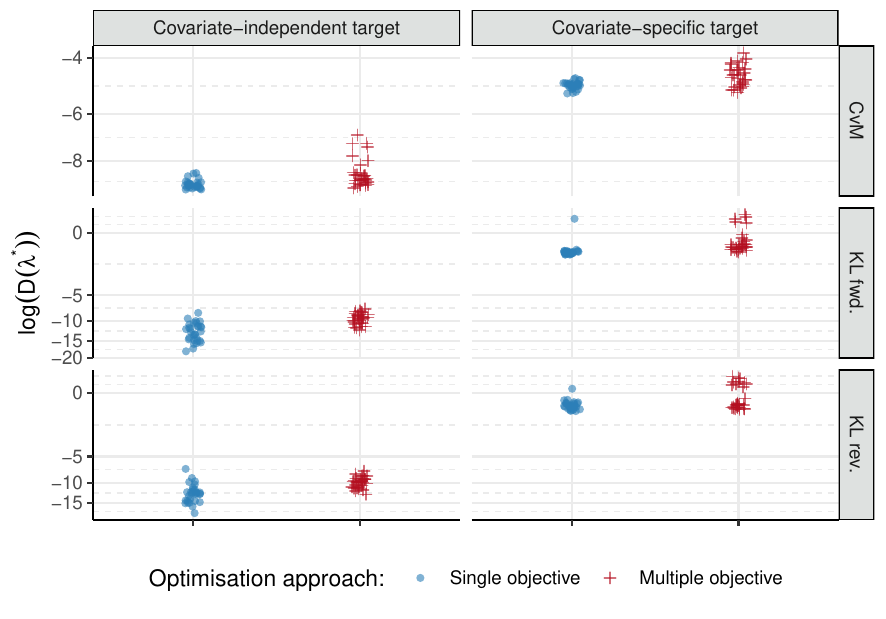} 

}

\caption{Final predictive discrepancy $\log(D(\lambda^{*}))$, or $\log(D(\lambda^{*} \mid \boldsymbol{X}))$ for the covariate-specific target. The multiple objective optimisation approach uses the optimal values $\kappa^{*}$ listen in Table \ref{tab:optimum-kappa-values} of this supplement. Horizontal jitter has been applied for readability.}\label{fig:discrep_at_optima}
\end{figure}

Figure \ref{fig:discrep_at_optima} displays the value of the discrepancy
(CvM or KL, as appropriate) at the optima located by the multi-stage
optimisation process. The optima are not comparable across targets,
discrepancies (the KL-divergence is not a distance metric), and
optimisation approaches, however for specific choices of these we can
asses the variability across replicates, to eliminate
incomplete-optimisation as a possible source of
non-replicability/non-faithfulness. There is universally additional
noise in the multiple objective approach, which is expected, and there
is some slightly bi-modality in both KL-based discrepancies for the
covariate-specific target. Overall, the multi-stage optimiser seems to
consistently locate acceptable optima.

\hypertarget{further-assessing-faithfulness}{%
\subsection{Further assessing
faithfulness}\label{further-assessing-faithfulness}}

After asserting that optima are consistently found by our multi-stage
optimisation process, we assess faithfulness by inspecting Figure
\ref{fig:pop_target_discreps} (in the main text) and Figure
\ref{fig:cov_target_discreps} in this supplement, which display the
prior predictive and covariate-(in)dependent target distributions. All
single-objective approaches are more faithful than their multi-objective
counterparts, which is expected given we sacrifice some amount of
faithfulness for variability in \(\theta\) when using the
multi-objective approach. Of the discrepancies, for the
covariate-independent target (Figure \ref{fig:pop_target_discreps}, main
text), the Cramér-von Mises discrepancy seems most faithful and
replicable. For the covariate-specific target (Figure
\ref{fig:cov_target_discreps}, this supplement), all discrepancies
result in similarly faithful priors and prior predictive distributions.
Both the single and multi-objective approaches struggle to match the
prior predictive distribution at all ages, with consistently poorer
faithfulness for \(X_{1} = 2\). Empirically, it does not seem possible
to match all four margins of the supplied target prior predictive
distributions simultaneously, given the mathematical structure of the
model. Lastly, because \(\tp(Y \mid X_{1} = 2)\) is substantially
narrower than the other targets, it is optimal, under the Cramér-Von
Mises discrepancy and the forward KL, to select wider priors better
matching the older age target distributions. The reverse KL is more
concentrated than both the Cramér-von Mises and the forward KL, which is
a known property of the KL in this direction
\citep{minka_Divergence_2005}. This concentration is also visible when
inspecting the prior predictives for the conditional mean,
\(\pd(h(t; \theta) \mid \lambda^{*})\), displayed in Figure
\ref{fig:regression_prior_pred} of the main text.

\hypertarget{further-details-of-assessing-prior-replicability-uniqueness-and-differences-between-kl-and-cvm-discrepancies}{%
\subsection{Further details of assessing prior replicability,
uniqueness, and differences between KL and CvM
discrepancies}\label{further-details-of-assessing-prior-replicability-uniqueness-and-differences-between-kl-and-cvm-discrepancies}}

\begin{figure}

{\centering \includegraphics{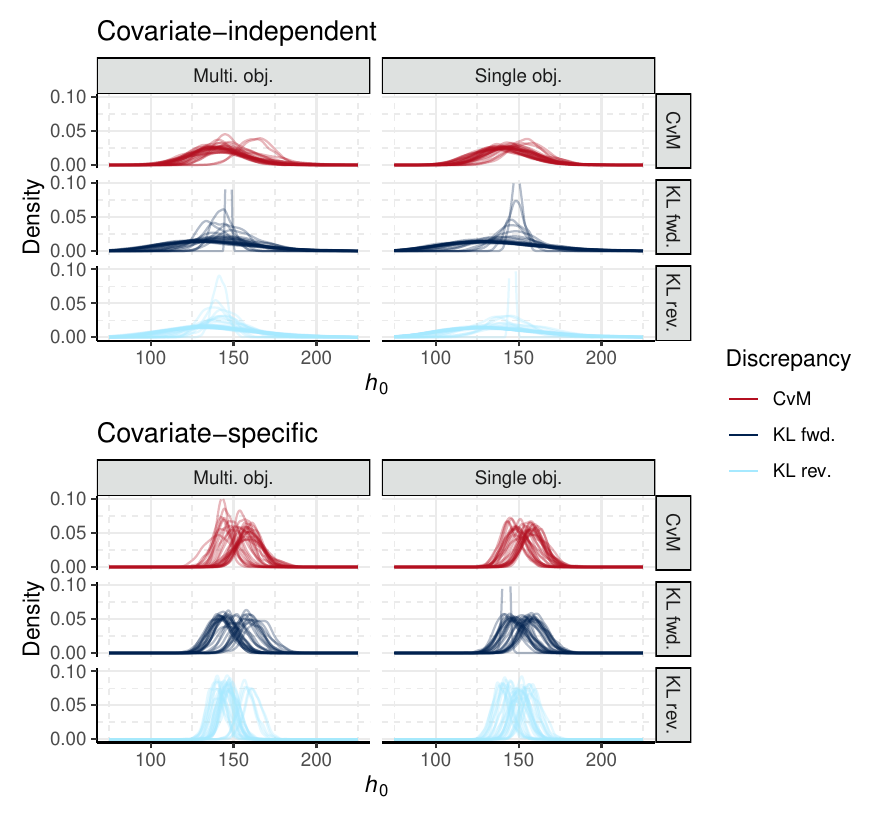} 

}

\caption{The marginal prior $\pd(h_{0} \mid \lambda^{*})$ translated from both the covariate-independent and covariate-specific targets, for all discrepancies considered in this example. Multi-objective priors are chosen using the relevant value of $\kappa^{*}$ in Table \ref{tab:optimum-kappa-values}.}\label{fig:theta_h0_across_discrep}
\end{figure}

We assess replicability and uniqueness in \(\theta\) by inspecting
\(\pd(h_{0} \mid \lambda^{*})\) displayed in Figure
\ref{fig:theta_h0_across_discrep}. All target and discrepancy
combinations seem to provide broadly replicable results when inspecting
\(h_{0}\), which agrees with the assessed optimisation convergence in
Figure \ref{fig:discrep_at_optima}. However uniqueness remains an
unsolved challenge, particularly for the covariate-specific target,
where two distinct modes are visible across both single and
multi-objective settings for all discrepancies. We highlight, for the
covariate-specific target, the very similar marginal priors found using
either the Cramér-von Mises, forward KL, or reverse KL discrepancy. This
indicates, for at least this parameter and target, that the optimal
prior is insensitive to the choice of discrepancy. We also observe wider
priors for \(h_{0}\) in the covariate-independent setting using the
KL-based discrepancies, which further explains the implausibly flat (a
priori) growth curves visible in Figure \ref{fig:regression_prior_pred}
of the main text.

\hypertarget{full-marginal-prior-and-posterior-comparison-plots}{%
\subsection{Full marginal prior and posterior comparison
plots}\label{full-marginal-prior-and-posterior-comparison-plots}}

Figures \ref{fig:pb_pop_prior_post_compare} and
\ref{fig:pb_cov_prior_post_compare} are extended versions of Figure
\ref{fig:small_cov_prior_post} in the main text, and display the prior
and posterior estimates for all the parameters in \(\theta\). Note that
results here are limited to the priors, and corresponding posteriors,
obtained using the Cramér-Von Mises discrepancy. Consistency and
uniqueness remain, evidently, challenging and as yet unobtainable.

\begin{landscape}
\begin{figure}

{\centering \includegraphics[width=0.8\linewidth]{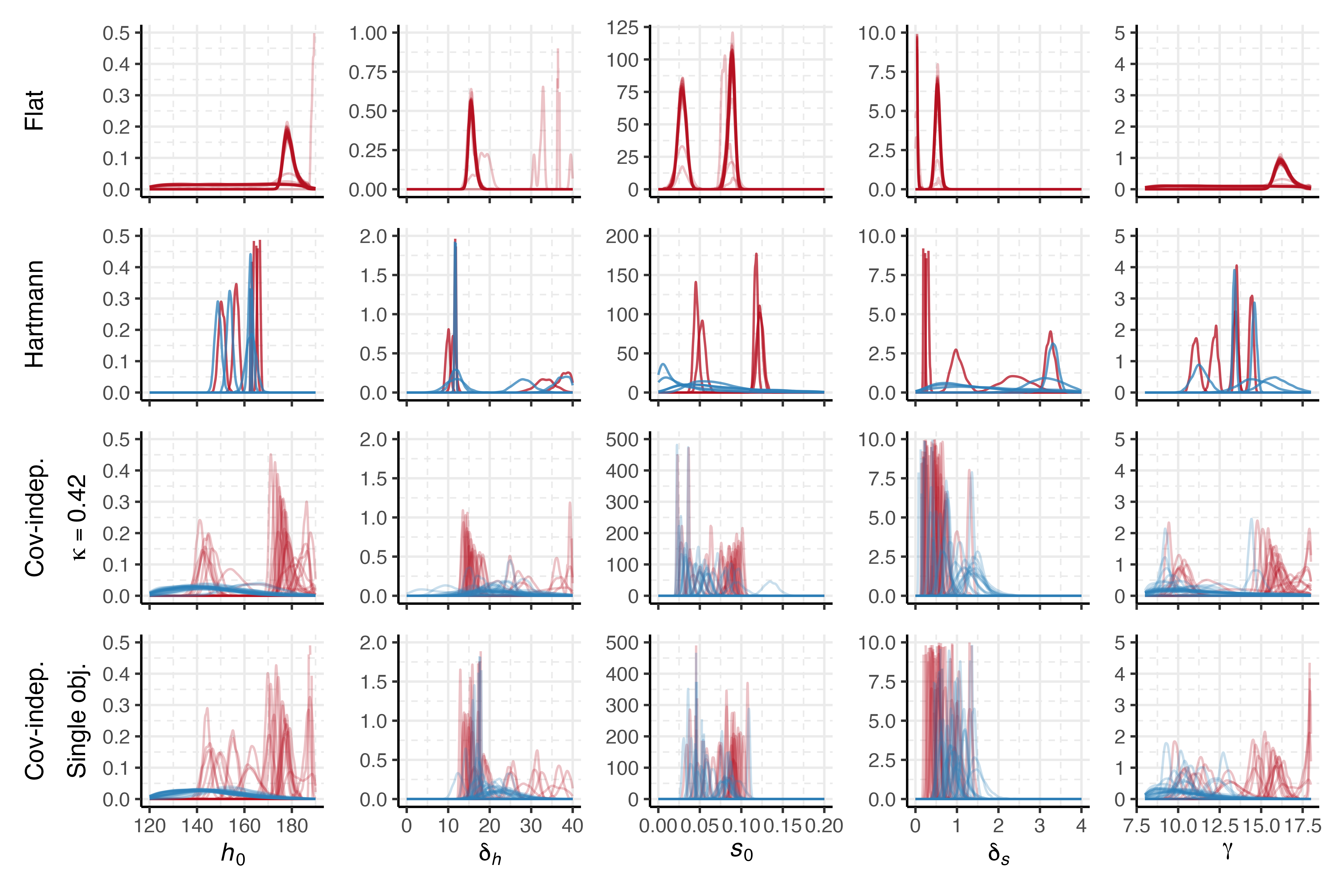} 

}

\caption{A comparison of the priors (\textcolor{mymidblue}{blue}) produced by our method using the covariate-independent marginal target (bottom two rows, Cramér-von Mises discrepancy only); and Hartmann et al. (2020) (second row), with no prior displayed for the flat prior scenario. The corresponding posteriors (\textcolor{myredhighlight}{red}) for individual $n = 26$ under each of these priors are displayed as dashed lines. Note that y-axes change within columns and are limited to values that clip some of the priors/posteriors for readability.}\label{fig:pb_pop_prior_post_compare}
\end{figure}

\begin{figure}

{\centering \includegraphics[width=0.8\linewidth]{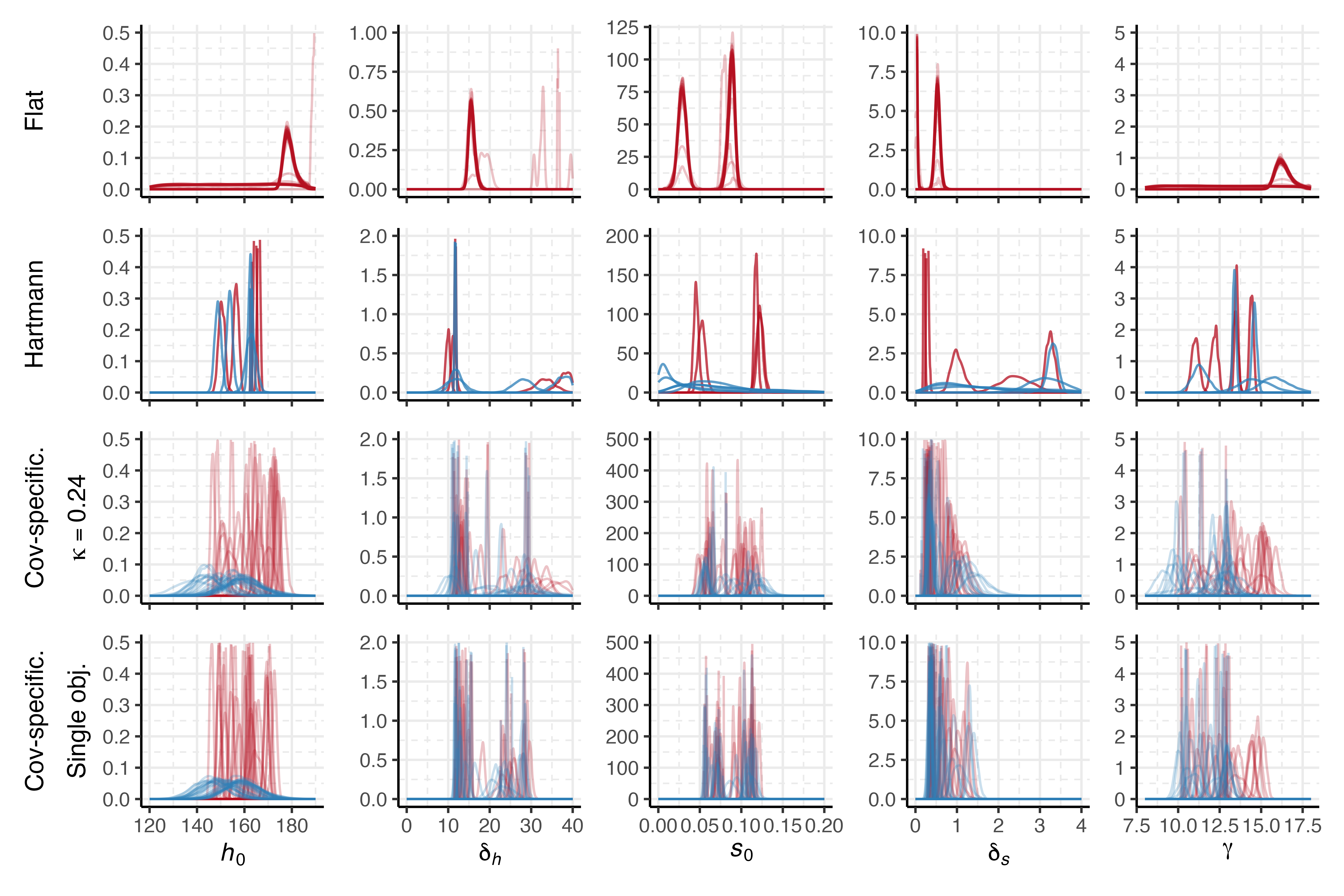} 

}

\caption{Otherwise identical to Figure \ref{fig:pb_pop_prior_post_compare} but the bottom two rows display the results obtained using the covariate-specific target.}\label{fig:pb_cov_prior_post_compare}
\end{figure}
\end{landscape}

\FloatBarrier

\bibliographystyle{agsm}
\bibliography{bibliography/prior-setting.bib}

\end{document}